\PassOptionsToPackage{numbers,sort&compress}{natbib}
\documentclass{article}
\usepackage[preprint]{neurips_2023}


\usepackage{xspace}

\usepackage{graphicx}
\usepackage{subcaption}
\usepackage{booktabs}
\usepackage{multirow}
\usepackage{makecell}

\usepackage{amsmath}
\usepackage{amssymb}

\usepackage[
    breaklinks,
    colorlinks=true,
    linkcolor=blue,
    citecolor=blue,
    urlcolor=blue
]{hyperref}
\usepackage{cleveref}

\usepackage{url}

\title{A Lensless Polarization Camera}

\author{
Noa Kraicer$^{1,\dagger}$,
Shay Elmalem$^{1,\dagger}$\thanks{Current affiliation: Department of Physics of Complex Systems, Faculty of Physics, Weizmann Institute of Science, Rehovot, Israel. The contribution to this work was made while affiliated with Tel Aviv University.},
Erez Yosef$^{1}$,
Hani Barhum$^{1,2}$,
Raja Giryes$^{1}$
}

\date{}

\begin{document}

\maketitle

\begin{center}
\small
$^1$ School of Electrical Engineering, The Iby and Aladar Fleischman Faculty of Engineering,
Tel Aviv University, Tel Aviv, Israel\\
$^2$ Triangle Regional Research and Development Center, Kfar Qara' 3007500, Israel\\[0.5em]
$^\dagger$ Equal contribution.
\end{center}

\begin{abstract}
Polarization imaging is a technique that creates a pixel map of the polarization state in a scene. Although invisible to the human eye, polarization can assist various sensing and computer vision tasks. Existing polarization cameras use spatial or temporal multiplexing, which increases the camera volume, weight, cost, or all of the above. Recent lensless imaging approaches, such as DiffuserCam, have demonstrated that compact imaging systems can be realized by replacing the lens with a coding element and performing computational reconstruction.
In this work, we propose a compact lensless polarization camera composed of a diffuser and a simple striped polarization mask. By combining this optical design with a reconstruction algorithm that explicitly models the polarization-encoded lensless measurements, four linear polarization images are recovered from a single snapshot.
Our results demonstrate the potential of lensless approaches for polarization imaging and reveal the physical factors that govern reconstruction quality, guiding the development of high-quality practical systems.\end{abstract}

\section{Introduction}
\label{sec:intro}

Polarization is a fundamental property of light, defined by the relative orientation and phase of the electric field components. 
While conventional imaging systems primarily measure intensity and spectral content, polarization provides complementary information about scene structure, material properties, and light–matter interactions.
Mapping the polarization state of a scene has enabled a wide range of sensing and computer vision applications, including visibility enhancement, material classification, and strain analysis in transparent media \cite{goldstein2017polarized,tyo2006review}.

State-of-the-art polarization cameras primarily utilize division-of-focal-plane (DoFP) or sequential filtering. Commercial DoFP sensors \cite{sonyPol} enable snapshot capture but sacrifice spatial resolution, requiring complex joint-demosaicing and super-resolution to mitigate artifacts \cite{zhou2025pidsr}. 
Beyond conventional micro-polarizer arrays, recent integrated snapshot designs based on metasurfaces \cite{huang2023high, li2025flat} and stacked architectures \cite{sasagawa2022polarization} have pushed chip-level integration, but still rely on intricate and high-precision fabrication processes. 
By contrast, integrated sequential polarization architectures \cite{liu2021high} preserve full spatial resolution, but require multiple exposures and therefore cannot capture dynamic scenes in a single shot.

As polarization imaging applications continue to expand, there is a growing need for a simple snapshot polarization camera.
In particular, applications such as compact microscopy and endoscopy, which may benefit from polarization imaging \cite{polMicroscopy,polImg_medical,polTissueRec}, impose strict constraints on system size and weight, often precluding the use of complex optics or even lenses \cite{WallerMiniScope}.
These constraints motivate the exploration of lensless imaging solutions.

Lensless imaging offers a hybrid optical-computational approach, where a simple optical element encodes the scene onto the sensor, and a post-processing algorithm reconstructs the image \cite{boominathan2021recent}.
Such systems have demonstrated the ability to capture not only two-dimensional intensity images \cite{flatCam}, but also additional modalities including depth \cite{DiffuserCam, PhlatCam}, temporal information \cite{WallerRlngShtr}, and spectrum \cite{diffSpec}, while significantly reducing system size and optical complexity.

\begin{figure}[t]
\begin{center}
\includegraphics[width=\linewidth]{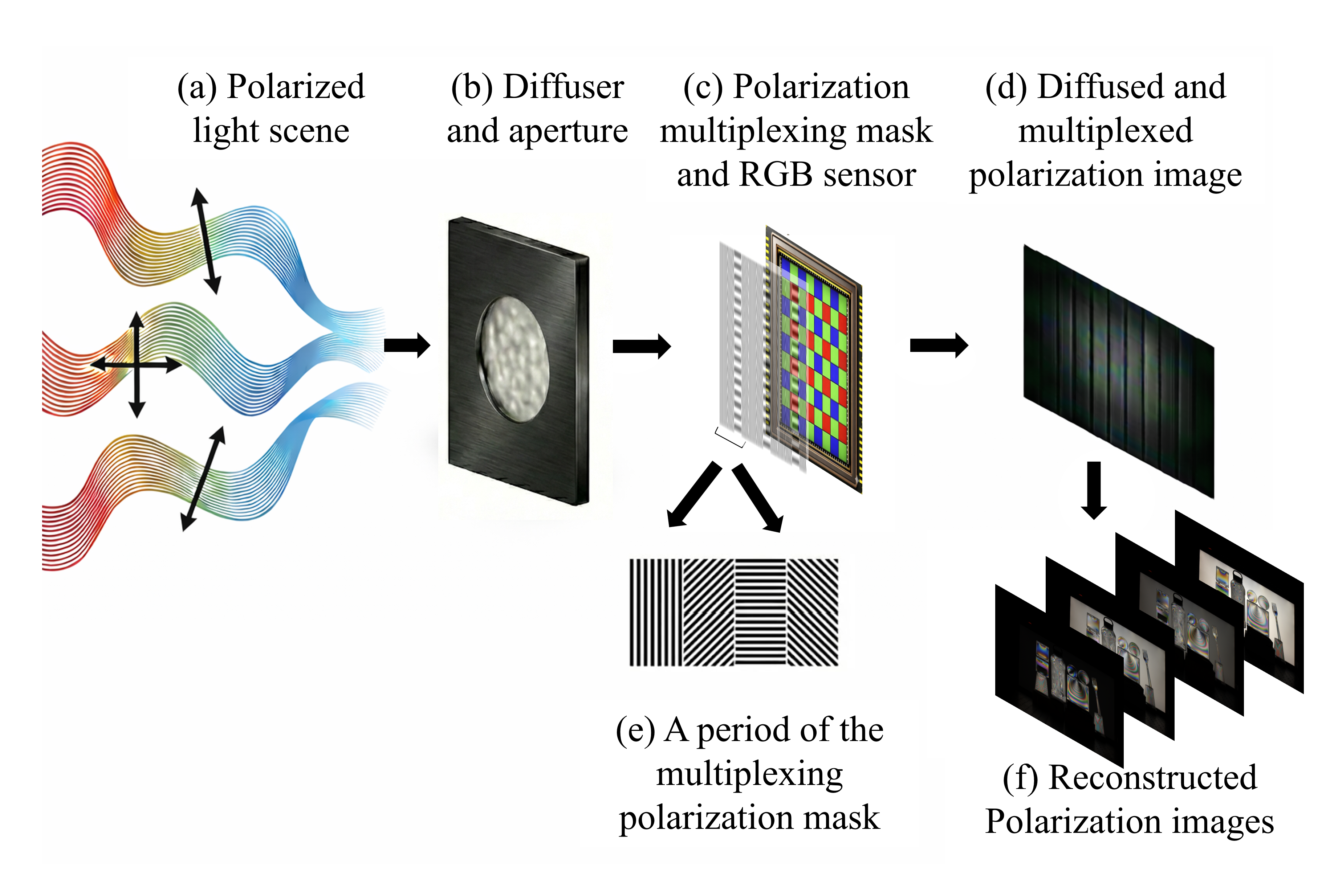}
\caption{{\bf Overview of the proposed lensless polarization camera} Given (a) a scene with polarized light, the imaging is performed using (b) a random diffuser. Polarization multiplexing is achieved using (c) a mask located on the sensor plane, where (e)  demonstrates a single period of the polarization multiplexing pattern, encoding linear polarization orientations ($0^\circ,45^\circ,90^\circ$ and  $135^\circ$). This imaging scheme results in (d) a diffused and multiplexed image. Using the diffuser PSF and polarization mask characteristics, the reconstruction algorithm recovers (f) polarization intensity images of the scene.}
\label{fig:polCam_system_diagram}
\end{center}
\end{figure}

Motivated by these advantages, we introduce a lensless polarization camera (see \Cref{fig:polCam_system_diagram}) that expands the idea briefly presented in  \cite{elmalem2021lensless}.
The proposed system is compact and low-cost, and enables single-shot recovery of polarization images suitable for downstream tasks, such as polarization-based strain analysis.
Our proposed camera is based on three main components:
(i) a random diffuser, similar to that used in DiffuserCam \cite{DiffuserCam}, whose point spread function (PSF) contains many narrow features distributed over almost the entire sensor area, enabling image reconstruction from a partially sampled image;
(ii) a simple striped polarization mask in which adjacent stripes have different linear polarization orientations (e.g., $0^\circ,45^\circ,90^\circ,135^\circ$), so that different sensor pixels measure different polarization components in a single exposure (i.e., polarization multiplexing); and
(iii) a physics-based reconstruction algorithm that explicitly models the system forward model, including the diffuser PSF and the polarization mask.

We evaluate the approach using matched forward-model simulations and real-world experiments with a laboratory prototype. 
While the key contribution of this work is the demonstration of a lensless polarization imaging system, we add a systematic analysis of the impact of model mismatch on reconstruction performance, to which lensless imaging systems are particularly sensitive. 
By isolating mask-induced model errors under controlled conditions, we identify mask mismatch as a dominant limiting factor in diffuser-based lensless polarization imaging. 
Importantly, the strong performance observed under matched-model simulations indicates that these limitations are not fundamental, but arise from prototype-level physical constraints, motivating tighter mask–sensor integration in future systems.

The rest of the paper is organized as follows: 
\Cref{sec:prevWork} provides a brief background on polarization imaging, existing lensless imaging approaches, and model mismatch in lensless imaging. The lensless polarization camera is presented in \Cref{sec:sys}, and its experimental performance is demonstrated and analyzed in \Cref{sec:exp}. \Cref{sec:discussion,sec:conclusion} discuss the results and conclude the paper.

\section{Background and Related Work}
\label{sec:prevWork}
\subsection{Polarization Imaging}
Polarization imaging refers to the pixelwise mapping of the polarization state in a scene. 
It can be fully described using the four Stokes parameters \cite{stokesParamPaper}:
\begin{equation}
\begin{cases}
    S_0 = I \\
    S_1 = Ip cos(2\psi)cos(2\xi) \\
    S_2 = Ip sin(2\psi)cos(2\xi) \\
    S_3 = Ip sin(2\xi),   
\end{cases}
\label{eq:stokesParams}
\end{equation}
where $S_i$ is the $i$-th Stokes parameter, $I$ is the total intensity, $p$ is the degree of polarization and $\psi,\xi$ are the angles of the polarization point on the Poincar\'e sphere \cite{poincare1892theorie}, indicating the angle and ellipticity of the polarization state. These parameters can be estimated for every point in the scene using several intensity images taken under different polarization states \cite{polImagOld}, which we refer to hereafter as \textit{polarization sub-images}:

\begin{equation}
\begin{cases}
    S_0 = I_{0} + I_{90} \\
    S_1 = I_{0} - I_{90} \\
    S_2 = I_{45} - I_{135} \\
    S_3 = I_{RCP} - I_{LCP}, 
\end{cases}
\label{eq:stokesImag}
\end{equation}
where $I_{\theta}$ denotes the intensity sub-image in linear polarization oriented at angle $\theta^\circ$ or in right/left circular polarization (RCP/LCP). 

Although polarization imaging was introduced decades ago \cite{polImagOld}, it was not widely adopted for many years.
Recently, advances in snapshot polarization sensing have renewed interest in polarization imaging across both research and applications.
These include commercial DoFP polarization cameras based on on-chip polarization sensor technology \cite{sonyPol,lucidPolCam}, as well as emerging compact full-Stokes designs based on metasurfaces and stacked or integrated polarization architectures  \cite{capassoPolCam,sasagawa2022polarization,li2025flat}.
The unique properties revealed by polarization imaging have enabled a wide range of applications, including imaging in extreme or scattering environments \cite{underwaterPolImg,fogPolImg,li2025robust}, biological and medical imaging \cite{polMicroscopy,polImg_medical,polTissueRec,guan2025review}, computer vision tasks such as shape and material estimation \cite{3dPolImg,deepSfp}, remote sensing \cite{remoteSensingPolImg}, and astronomical observations including exoplanet detection \cite{zimpol}.
This proliferation has also stimulated algorithmic work on polarization image processing, including demosaicing for polarization mosaics \cite{polDem_CNN,KaustPolDem,TokyoTech_polDem,pistellato2022deep,zhou2025pidsr,zheng2024color}, as well as dedicated denoising and deblurring methods \cite{polDenoise_PCA,polDenoiseLearn,liu2024review,zhou2025learning}.

\subsection{Lensless Imaging}
Miniaturization is a serious challenge in imaging system design. In the last two decades, the possibility of reducing the space and weight of lens-based cameras while keeping high imaging performance has been drastically improved (mostly thanks to the smartphone industry). However, extreme imaging applications (like in-vivo microscopy or endoscopy) might prevent the use of conventional lenses \cite{Katz_lenslessFiberBundle, Katz_lenslessEndoscopy_18, WallerMiniScope}. To address this challenge, several lensless imaging methods were introduced recently \cite{DiffuserCam,flatCam, PhlatCam,lenslessMenon,lenslessBarbasthatis, Katz_lenslessFiberBundle, Katz_lenslessEndoscopy_18}. These methods are based on a conventional image sensor and a mask that replaces the lens in the sense that it has some sort of PSF. Such a PSF is generally poor, and its generated image is practically random. However, it has unique properties that enable a post-capture reconstruction of the image, achieving high-quality lensless imaging reconstruction performance. The mask can modulate either amplitude \cite{flatCam} or phase \cite{PhlatCam, DiffuserCam,lenslessBarbasthatis}, be designed \cite{flatCam, PhlatCam, yosef2024difuzcam} or random \cite{DiffuserCam}, and in some cases, even a bare sensor \cite{lenslessMenon} or fiber bundle \cite{Katz_lenslessFiberBundle, Katz_lenslessEndoscopy_18} can suffice. A comprehensive review of lensless imaging systems can be found in \cite{boominathan2021recent}. In addition to conventional 2D imaging, lensless cameras can capture additional modalities, e.g., depth \cite{DiffuserCam,flatCam, PhlatCam}, time \cite{WallerRlngShtr}, or spectrum \cite{diffSpec}. In this work, we leverage the unique properties of lensless imaging together with polarization imaging to design a simple lensless polarization camera.

Recent work has explored lensless acquisition of polarization and additional modalities using diffuser or phase-mask optical encoding.
Baek et al.~\cite{baek2022lensless} demonstrated a single-shot full-Stokes lensless polarization camera based on a phase mask and a polarization-encoded aperture, showing recovery of both linear and circular polarization components.
Zheng et al.~\cite{zheng2023spectral} proposed joint spectral and polarization imaging in a lensless diffuser camera using space-division multiplexing, demonstrated only in simulation, and Diffuser-mCam~\cite{zheng2025exploiting} further extended this paradigm to simultaneous spectral, polarization, and temporal imaging via aggressive compressed sensing.

While these approaches emphasize modality multiplexing and polarization completeness, they rely on spatially partitioned apertures or aggressive compression to enable joint recovery of multiple modalities.
In particular, in \cite{baek2022lensless}, the polarization-dependent forward operators are approximated by spatially partitioning a single measured PSF according to the aperture layout.
In Diffuser-mCam~\cite{zheng2025exploiting}, polarization- and wavelength-dependent PSFs are calibrated explicitly, but reconstruction is performed after strong spatial downsampling and the forward model is implemented as a generic linear operator, reflecting the assumed limited shift-invariance of the strongly scattering diffuser.
As a result, the effective spatial resolution per polarization component is reduced, and reconstruction can become more sensitive to forward-model mismatch.

In contrast, our work focuses on linear polarization imaging and leverages the spatial multiplexing induced by a diffuser to enable image-plane polarization multiplexing with a striped polarization mask. 
We operate in a regime in which the diffuser response can be approximated as shift-invariant, resulting in a forward model with a convolutional diffuser operator defined on the native spatial grid.
This design avoids polarization-dependent aperture partitioning and enables a more interpretable forward model.
Beyond demonstrating snapshot recovery of four linear polarization sub-images, we further analyze sensitivity to polarization-mask mismatch, which we show to be a dominant factor limiting the practical performance of both our proposed method and other approaches.

\subsection{Model Mismatch in Lensless Imaging}

Non-blind image reconstruction is traditionally formulated as a trade-off between model-based inversion and prior compliance \cite{RUDIN92Nonlinear,bertero2021introduction}. Although model-based approaches are theoretically well grounded, their performance depends on the accuracy of the assumed forward model. In practical imaging systems, unavoidable deviations between the physical acquisition process and the assumed model, commonly referred to as \emph{model mismatch}, can degrade reconstruction quality and lead to structured artifacts or bias in the recovered images, particularly in ill-posed inverse problems \cite{antun2020instabilities,burger2019convergence}.

In lensless imaging systems, model mismatch is often more complex due to the simplified assumptions typically made in the forward model. The imaging process is commonly approximated as a linear, shift-invariant convolution with a measured PSF. In practice, deviations from this idealized convolution model arise from the physical measurement process and have been shown to affect reconstruction performance in real lensless systems \cite{DiffuserCam,zeng2021robust,cai2024phocolens}.

By adding a polarization mask to the diffuser, the forward model becomes more structured and admits additional sources of mismatch. These include boundary effects at polarization stripe interfaces, the distance between the polarization mask and the sensor, and fabrication-related imperfections. Such effects introduce structured deviations from the assumed forward model that are not well captured by simple additive noise assumptions.
Similar mask-sensor distance and induced cross-talk effects have been reported in a lensless snapshot hyperspectral camera \cite{raniwala2023improved}, where they lead to systematic deviations from the assumed forward model and motivate improved calibration.

Recent advances in learning-based and hybrid reconstruction methods have proposed various strategies to mitigate modeling inaccuracies by incorporating learned priors or implicit regularization for lensless imaging reconstruction \cite{ongie2020deep,DIP,wallerUntrained,qian2024robust,cai2024phocolens,zeng2021robust}. While these approaches can significantly improve robustness under certain conditions, their application to newly designed optical systems introduces additional challenges in practice.
In particular, for polarization lensless imaging systems, acquiring large and diverse training datasets that faithfully reflect the physical acquisition process is often impractical, as polarization states cannot be arbitrarily synthesized using standard computer-display setups \cite{yosef2024difuzcam,hung2025scalable,khan2020flatnet,bezzam2025towards,monakhova2019learned}. 
Moreover, variations in fabrication processes and system assembly can introduce larger inter-system differences than in more mature lensless architectures, further complicating transfer across designs.

\begin{figure}[t]
\centering
\makebox[0.92\linewidth]{%
\begin{subfigure}[t]{0.30\linewidth}
    \centering
    \includegraphics[height=0.15\textheight]{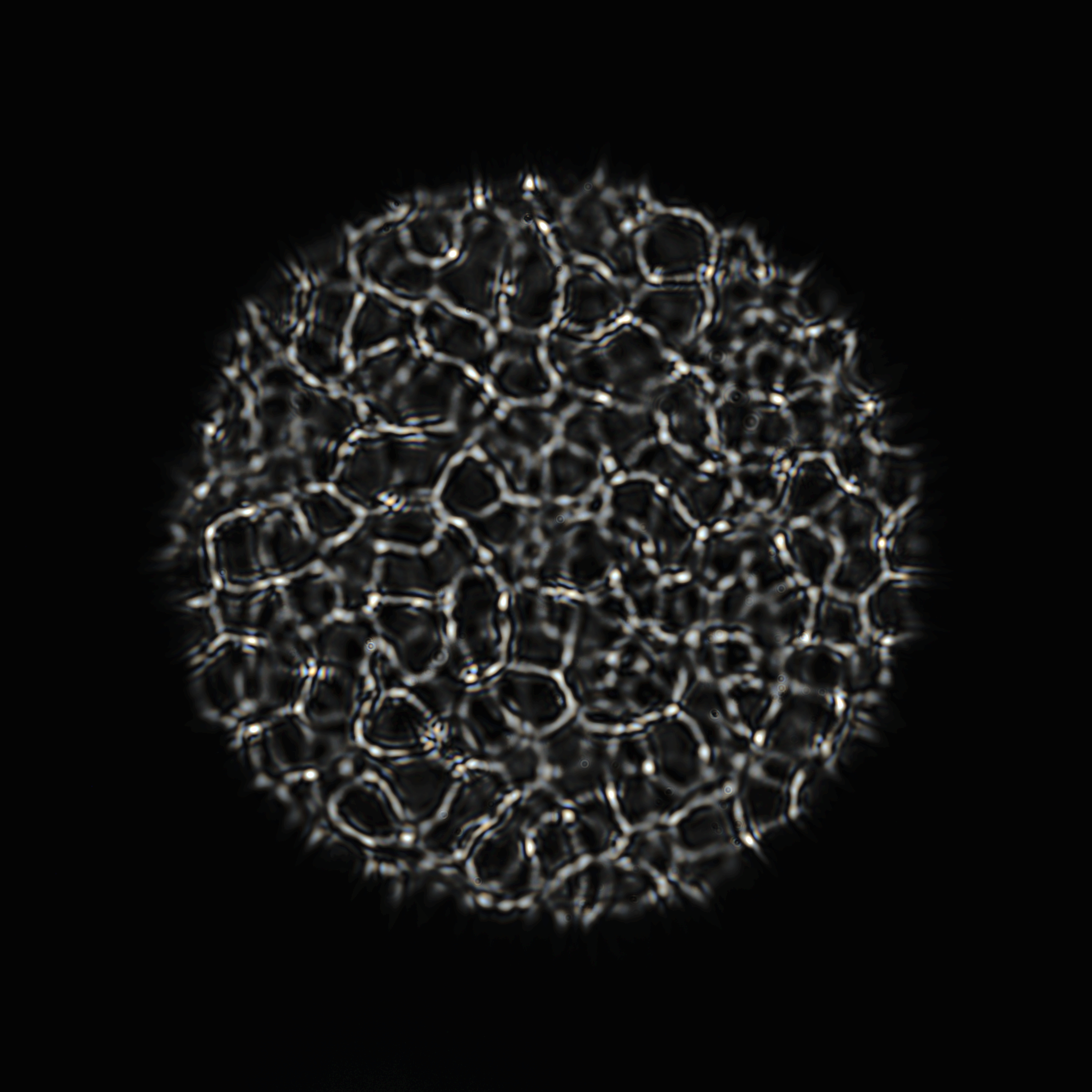}
    \subcaption{Diffuser PSF}
    \label{fig:PSF}
\end{subfigure}
\hfill
\begin{subfigure}[t]{0.25\linewidth}
    \centering
    \includegraphics[height=0.15\textheight,width=\linewidth]{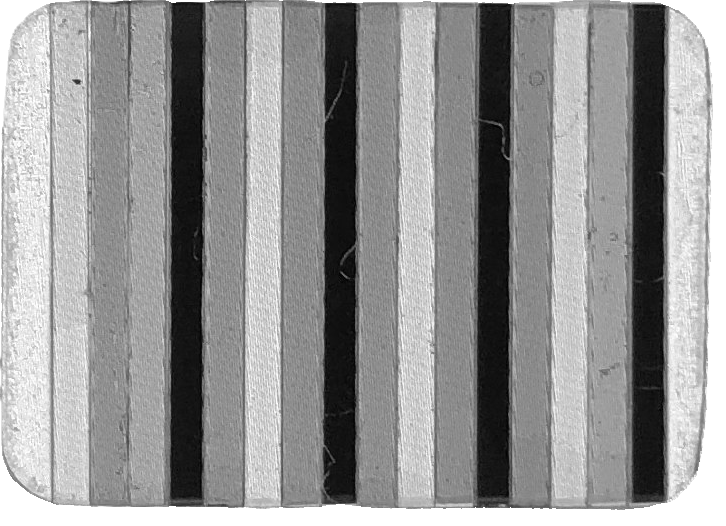}
    \subcaption{Polarization multiplexing mask}
    \label{fig:polfilt}
\end{subfigure}
\hfill
\begin{subfigure}[t]{0.30\linewidth}
    \centering
    \includegraphics[height=0.15\textheight]{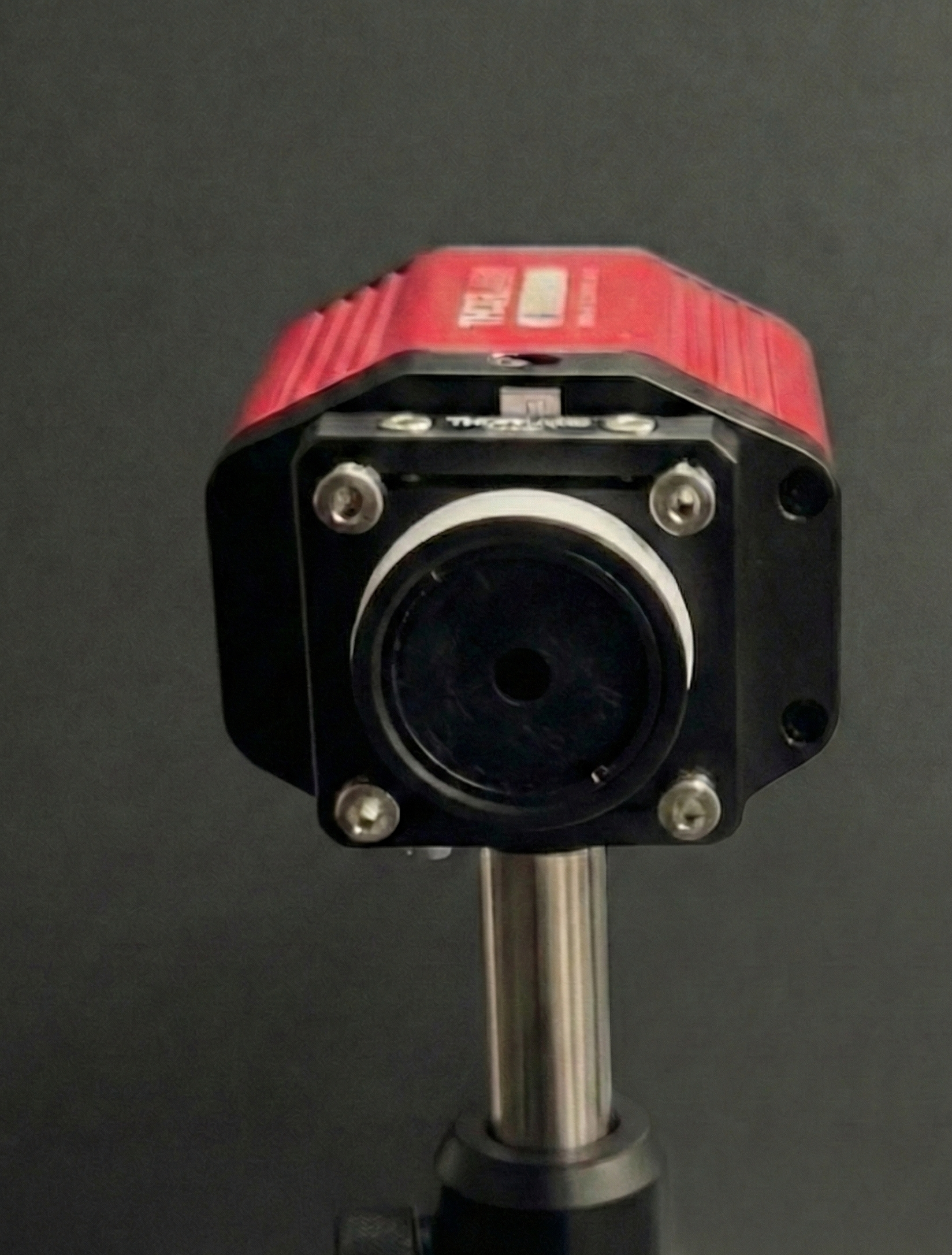}
    \subcaption{Lensless polarization camera}
    \label{fig:polcam}
\end{subfigure}
}
\caption{\textbf{Optical encoding elements.} 
(a) Diffuser PSF. The random structure of the diffuser acts as multiple randomly distributed impulse-like features. This produces a spatially extended PSF with many narrow features that enable image reconstruction. 
(b) Polarization mask. Leveraging the ability to reconstruct images from partial sampling of the diffuser-encoded measurement, the mask is composed of linear polarizer stripes at the required orientations. The mask response is measured under back-illumination using a polarized light source, enabling the different transmissions to be clearly visualized. 
(c) A prototype lensless polarization camera integrating the diffuser and polarization mask at the sensor.}

\label{fig:opticalElements}
\end{figure}

\section{Method}
\label{sec:sys}

We present a lensless polarization camera that combines a random diffuser for spatial encoding of the scene with a polarization multiplexing mask (\Cref{fig:opticalElements}), enabling acquisition of multiple polarization sub-images in a single shot. 

To reconstruct the polarization sub-images, we use a physics-based algorithm that relies on the resulting encoding. In the following subsections, we describe the diffuser-based imaging principle (\Cref{subsec:diffuser}), the polarization mask (\Cref{subsec:polarization}), the reconstruction algorithm (\Cref{subsec:reconstruction}), and the physical prototype implementation (\Cref{subsec:prototype_structure}).

\subsection{Diffuser-Based Imaging}
\label{subsec:diffuser}
Inspired by the DiffuserCam concept \cite{DiffuserCam}, the proposed lensless polarization camera is based on a random diffuser as the primary optical encoding element.
The diffuser spatially multiplexes the incoming light, such that the response of each scene point is spread over many sensor pixels, producing a spatially extended PSF that simultaneously exhibits many spatially localized, high-contrast, approximately impulse-like features (see \Cref{fig:PSF}).
This multiplexing property enables recovery of scene information from a subset of sensor pixels and allows additional modalities to be encoded through spatial modulation.

As a result, the recorded measurement can be interpreted as a linear superposition of multiple spatially shifted and encoded versions of the scene. 
This property has previously been exploited to encode and reconstruct additional modalities alongside spatial information, such as time and spectrum \cite{WallerRlngShtr,diffSpec}. In this work, we leverage the same principle to enable polarization reconstruction.
For simplicity, we restrict our system to a regime in which the diffuser PSF can be approximated as spatially shift-invariant, allowing a linear shift-invariant (LSI) forward model. This assumption is not fundamental to the proposed imaging scheme, and a shift-variant model could be employed with some modifications in the reconstruction method.

\subsection{Polarization Mask}
\label{subsec:polarization}

Building on the spatial multiplexing capability provided by diffuser encoding, we introduce a polarization multiplexing mask that enables the simultaneous acquisition of multiple polarization images within a single measurement. 

Polarization imaging requires overcoming an inherent hurdle: While polarization is a property of the electromagnetic wave’s complex amplitude, conventional image sensors measure only the light intensity and are inherently polarization-insensitive. Therefore, enabling polarization imaging requires incorporating polarization-sensitive optical elements and sampling intensity under multiple polarization states~\cite{polImagOld}. Full reconstruction of the Stokes parameters for each pixel requires six polarization sub-images, as presented in \Cref{eq:stokesImag}. Yet, linear-only polarization mapping can be achieved using only four sub-images. In this work, we limit the reconstruction to linear polarization mapping. However, extension to circular/elliptical polarization is possible. 

For linear polarization mapping, four polarization sub-images are required: $I_{\theta}$, where  $\theta\in \{0^\circ,45^\circ,90^\circ,135^\circ\}$. Note that while linear polarization mapping is also possible using three sub-images \cite{polImagOld}, the common practice is to use four, for more stable results \cite{sonyPol,salsaPolCam,capassoPolCam}. To avoid complex pixel-scale fabrication, and by leveraging the partial sampling capability of the diffuser-based imaging scheme, we employ a simple striped polarization mask (\Cref{fig:polmask}), with a periodic pattern of linear polarizer stripes in the required angles. Its structure enables partial sampling in each of the required polarization angles. The main trade-off in such a mask design is the individual stripe width, as narrower stripes perform a denser sampling, with the cost of a more complex fabrication. As very dense sampling is not critical, relatively wide stripes, which are easy to fabricate, can be used, as detailed in \Cref{subsec:prototype_structure}.

\subsection{Reconstruction Algorithm}
\label{subsec:reconstruction}
The goal of the reconstruction algorithm is to integrate the diffuser PSF imaging features and the polarization multiplexing mask structure into a joint reconstruction of the four polarization sub-images. 

\noindent\textbf{Forward model:}
The polarization-independent diffuser spatially encodes the incoming light according to its PSF.
The polarization mask, located after the diffuser, contains stripes of four fixed polarizer orientations. 
Each stripe transmits the component of the light aligned with its orientation according to the polarizer transmission response.

Let $\mathbf{x} \in \mathbb{R}^{H \times W \times C \times P}$ denote the intensities of the polarization-resolved scene, where $H$ and $W$ are the spatial dimensions, $C=3$ is the number of color channels, and $P=4$ corresponds to the polarization orientations $0^\circ,45^\circ,90^\circ,$ and $135^\circ$.
Each $\mathbf{x}_{c,p} \in \mathbb{R}^{H \times W}$ represents the polarization-resolved intensity component at orientation $p$ in color channel $c$.

The mask is represented by spatial maps $\mathbf{S}_p$ that encode the projection imposed by the orientation of the stripe.

For each color channel $c$, the measurement is:
\begin{equation}
\label{eq:forward_model}
\mathbf{y}_{c}
= \sum_{p=1}^{P}
\mathbf{S}_{p}\,\odot\,
\bigl(\mathbf{x}_{c,p} * \mathbf{k}_c\bigr),
\end{equation}

where $\mathbf{k}_c$ is the diffuser PSF for channel $c$, $\odot$ denotes element-wise multiplication, and $*$ denotes 2D convolution over the spatial dimensions.

\noindent\textbf{Inverse problem formulation:}
Given the forward model in \Cref{eq:forward_model}, the reconstruction problem can be written in operator form as
\(
\mathbf{y} = A\mathbf{x} + \mathbf{e},
\)
where \(A\) denotes the linear forward operator encoding convolution with the diffuser PSF and polarization multiplexing by the mask, and \(\mathbf{e}\) represents additive measurement noise.
Reconstruction is formulated as the minimization of a regularized least-squares objective,
\begin{equation}
\label{eq:costFunc}
\hat{\mathbf{x}} = \arg\min_{\mathbf{x}} 
\frac{1}{2\sigma_e^2} \left\| \mathbf{y} - A \mathbf{x} \right\|_2^2 + s(\mathbf{x}),
\end{equation}
where \(\sigma_e\) denotes the standard deviation of the measurement noise.
The first term enforces data fidelity, while \(s(\mathbf{x})\) is a regularization term promoting desirable image priors.

Directly minimizing \Cref{eq:costFunc} is computationally challenging due to the ill-posed nature of the inverse problem.
We therefore employ the Alternating Direction Method of Multipliers (ADMM) \cite{ADMM}, using variable splitting to decouple the data fidelity and regularization terms.
In our implementation, \(s(\mathbf{x})\) is chosen as a weighted anisotropic Total Variation (TV) regularizer \cite{kamilov2016parallel}.
Additional implementation details appear in the Supplement Document (\Cref{sec:appendix_physics_based}).

 \subsection{Prototype Design}
 \label{subsec:prototype_structure}
The prototype lensless polarization camera is based on a $0.5^\circ$ diffuser (Edmund Optics \#47-860) mounted on a 12.3MP, $3.45[\mu m]$ pixel pitch CMOS camera (Thorlabs CS126CU). The polarization mask is fabricated using a linear film polarizer (Thorlabs LPVISE2X2) cut to stripes of $\approx880\,\mu\mathrm{m}$ (equivalent to $\approx 256[pix]$) oriented in the required polarization angles. The stripes are then assembled in the required form, and the polarization mask is placed directly above the image sensor.
Due to the presence of the sensor cover glass, a finite distance of approximately $1.36\,\mathrm{mm}$ exists between the polarization mask and the sensor pixel plane. As a result, the effective polarization modulation deviates from the idealized mask assumed in the forward model in \Cref{eq:forward_model}.
The diffuser PSF, which is assumed to be polarization independent, is measured by imaging a point light source after removing the polarization mask (see \Cref{fig:PSF}). 
The polarization mask response is measured separately, without the diffuser, using a spatially uniform broadband light source and an external rotating linear polarizer. We measure the polarization mask response for each polarization orientation by setting the external polarizer to the corresponding angle (see \Cref{fig:polfilt} for the measured response at a single representative angle).
This modular prototype design allows independent measurement of the diffuser PSF and polarization mask response, while also exposing potential sources of polarization-mask mismatch arising from assembly and fabrication tolerances.
The measured diffuser PSF and polarization mask response are incorporated into the forward operator in \Cref{eq:forward_model} and are used for reconstructing the polarization images acquired with our prototype.

\section{Experimental Results}
\label{sec:exp}
We evaluate the proposed lensless polarization imaging framework under increasing levels of realism. 
We first establish a proof-of-concept under ideal (matched-model) conditions, and then study the impact of forward-model mismatch caused by polarization-mask deviations.
Specifically, we present fully matched-mask simulations, in which the same polarization mask is used for data generation and reconstruction, to define an upper bound on achievable performance (\Cref{subsec:matched}), followed by controlled polarization-mask mismatch experiments to assess robustness (\Cref{subsec:mismatch}), and finally real-world results obtained with a laboratory prototype to demonstrate practical performance (\Cref{subsec:real-world}).
These experiments show that high-quality reconstruction is achieved when the forward model matches the acquisition process, whereas mask mismatch causes noticeable degradation.
The remaining performance gap in the laboratory prototype reconstructions therefore arises from physical deviations of the fabricated polarization mask from the assumed model, dominated by the finite mask–sensor separation, with additional contribution from fabrication imperfections, highlighting the importance of tighter sensor–mask integration in future designs.

\begin{figure}[t!]
\def\resImgSz{0.20}
\centering
\newcommand{\rotateshift}{2.5cm}

\begin{tabular}{c c c c c}

        \rotatebox{90}{\parbox[c]{\rotateshift}{\centering reconstruction  }} &
\includegraphics[width=\resImgSz\columnwidth]{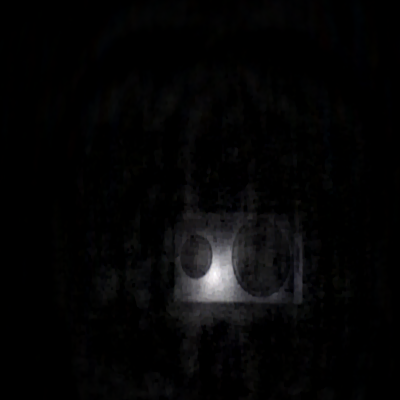} &
\includegraphics[width=\resImgSz\columnwidth]{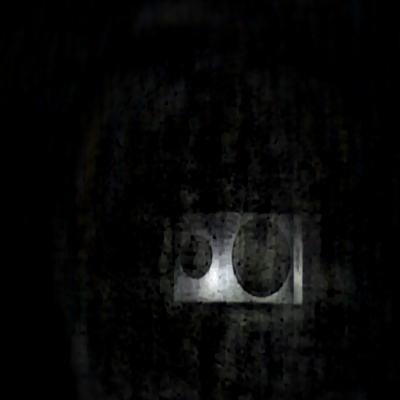} &
\includegraphics[width=\resImgSz\columnwidth]{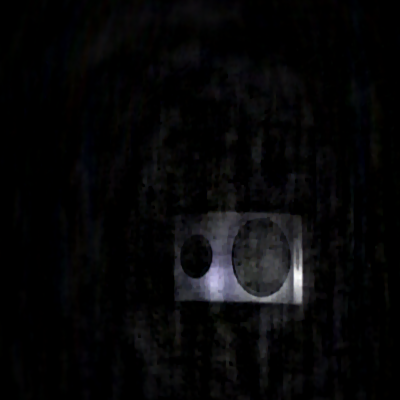} &
\includegraphics[width=\resImgSz\columnwidth]{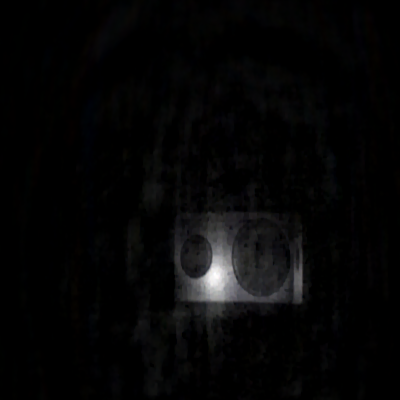} \\

        \rotatebox{90}{\parbox[c]{\rotateshift}{\centering reference  }} &
\includegraphics[width=\resImgSz\columnwidth]{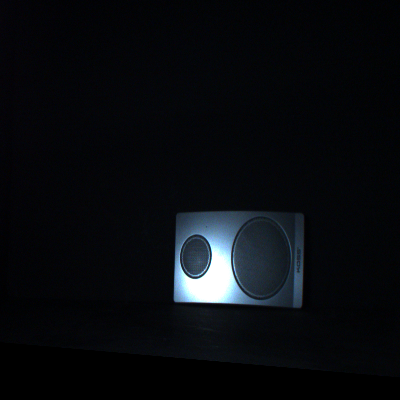} &
\includegraphics[width=\resImgSz\columnwidth]{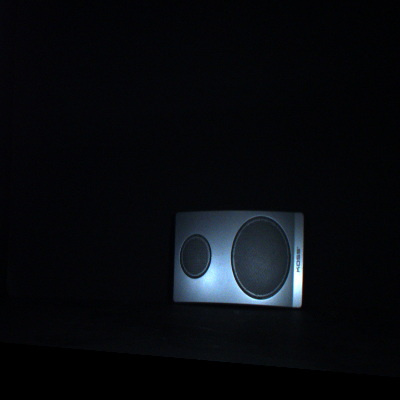} &
\includegraphics[width=\resImgSz\columnwidth]{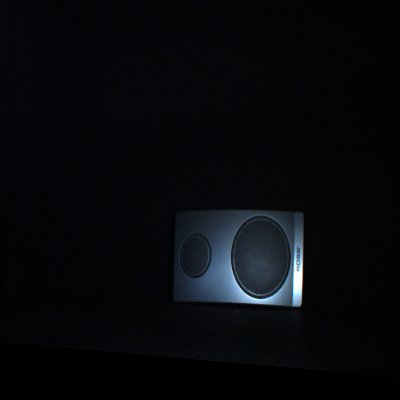} &
\includegraphics[width=\resImgSz\columnwidth]{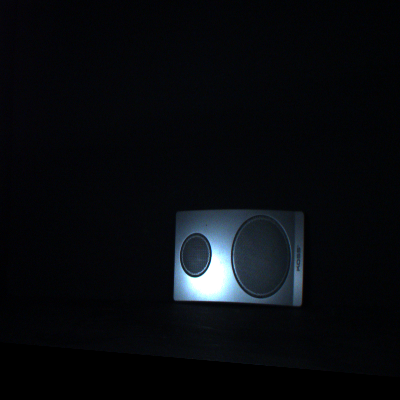} \\[1mm]

        \rotatebox{90}{\parbox[c]{\rotateshift}{\centering reconstruction  }} &
\includegraphics[width=\resImgSz\columnwidth]{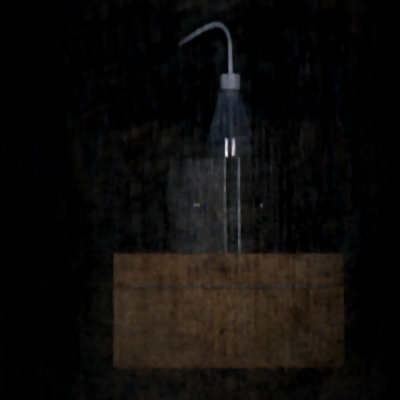} &
\includegraphics[width=\resImgSz\columnwidth]{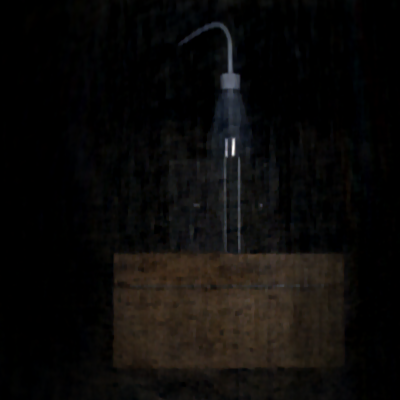} &
\includegraphics[width=\resImgSz\columnwidth]{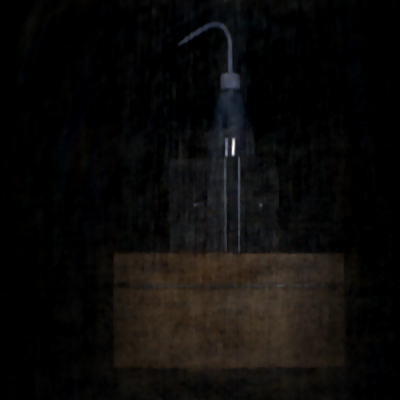} &
\includegraphics[width=\resImgSz\columnwidth]{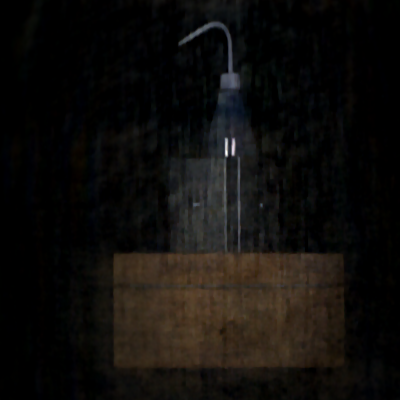} \\

        \rotatebox{90}{\parbox[c]{\rotateshift}{\centering reference  }} &
\includegraphics[width=\resImgSz\columnwidth]{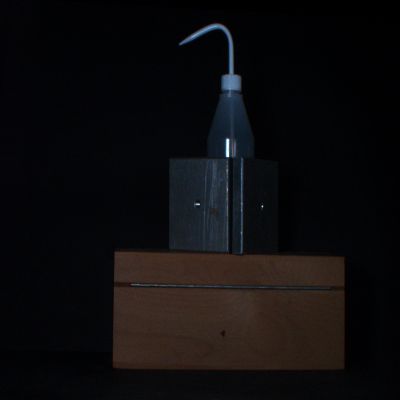} &
\includegraphics[width=\resImgSz\columnwidth]{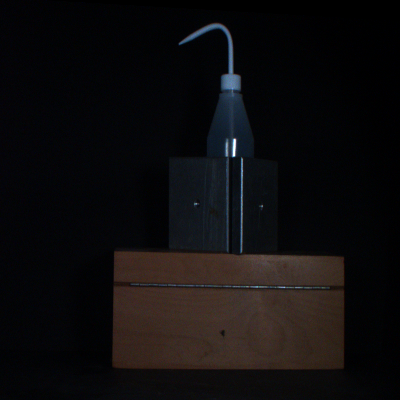} &
\includegraphics[width=\resImgSz\columnwidth]{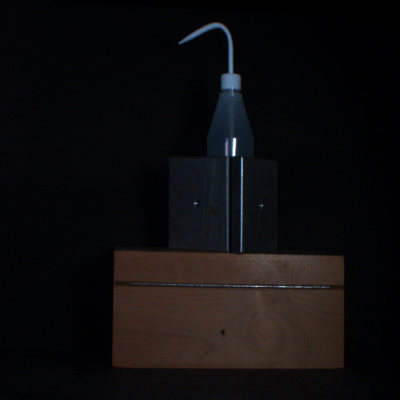} &
\includegraphics[width=\resImgSz\columnwidth]{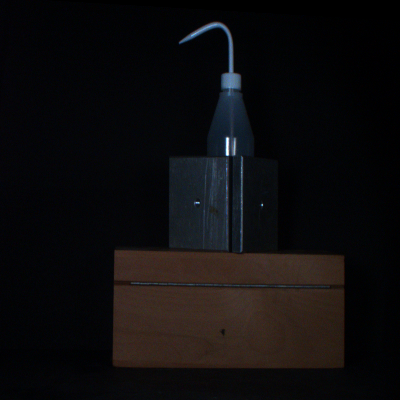} \\[1mm]

        \rotatebox{90}{\parbox[c]{\rotateshift}{\centering reconstruction  }} &
\includegraphics[width=\resImgSz\columnwidth]{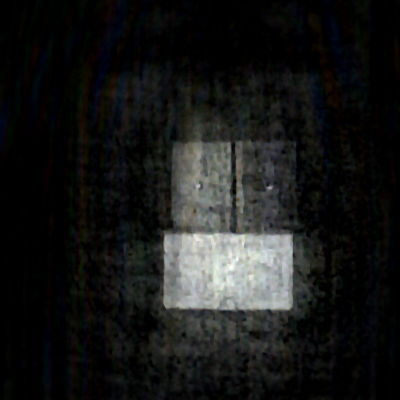} &
\includegraphics[width=\resImgSz\columnwidth]{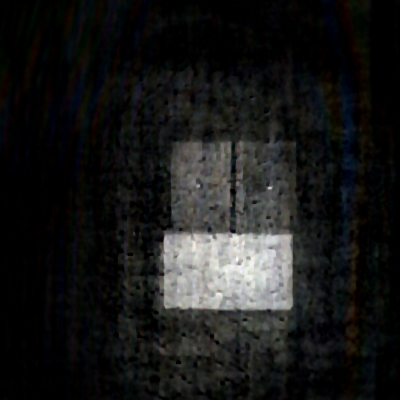} &
\includegraphics[width=\resImgSz\columnwidth]{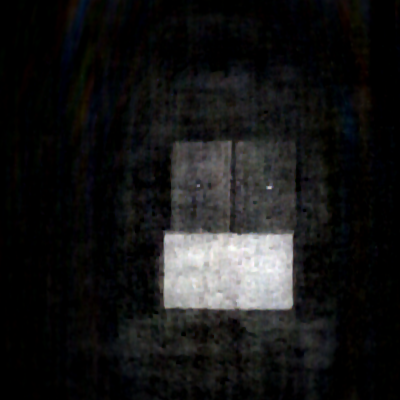} &
\includegraphics[width=\resImgSz\columnwidth]{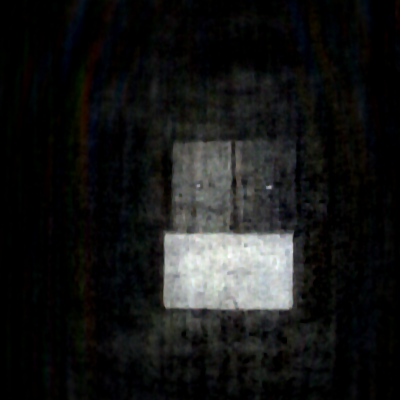} \\

        \rotatebox{90}{\parbox[c]{\rotateshift}{\centering reference  }} &
\includegraphics[width=\resImgSz\columnwidth]{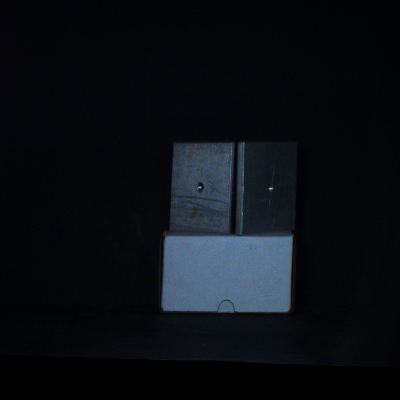} &
\includegraphics[width=\resImgSz\columnwidth]{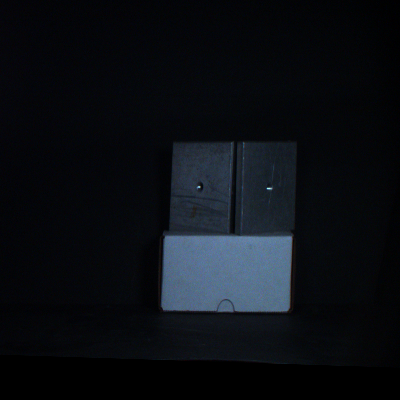} &
\includegraphics[width=\resImgSz\columnwidth]{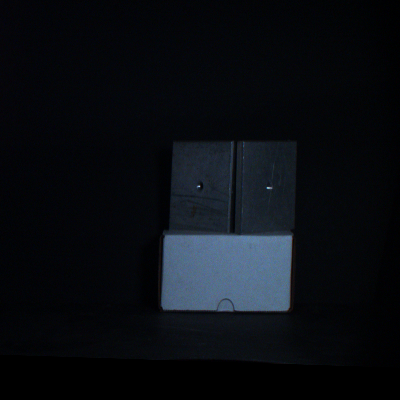} &
\includegraphics[width=\resImgSz\columnwidth]{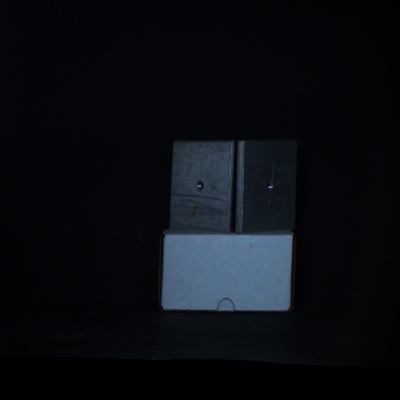} \\

&
\small{(a) $I_{0}$} &
\small{(b) $I_{45}$} &
\small{(c) $I_{90}$} &
\small{(d) $I_{135}$} \\

\end{tabular}

\vspace{-1mm}
\caption{
\textbf{Front illumination — matched-forward-model simulation.} For each scene, the first row shows the lensless polarization reconstruction with the same polarization mask for data generation and reconstruction, and the second row shows the corresponding spatially aligned reference images acquired using a lens-based camera with an external rotating polarizer.}
\label{fig:sim-front-ill}
\vspace{-2mm}
\end{figure}

\begin{figure}[t]
    \def\resImgSz{0.2}
	\centering
    \newcommand{\rotateshift}{2.5cm}
	\begin{tabular}{c c c c c}
	    \rotatebox{90}{\parbox[c]{\rotateshift}{\centering measured mask  }} &
        \includegraphics[width = \resImgSz\columnwidth]{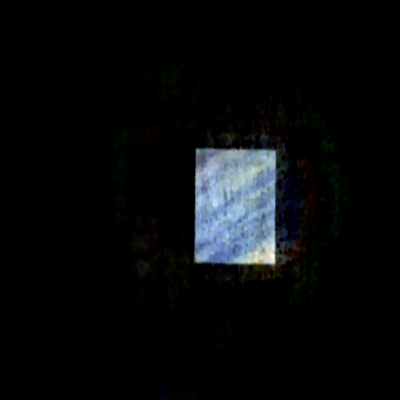} &
		\includegraphics[width = \resImgSz\columnwidth]{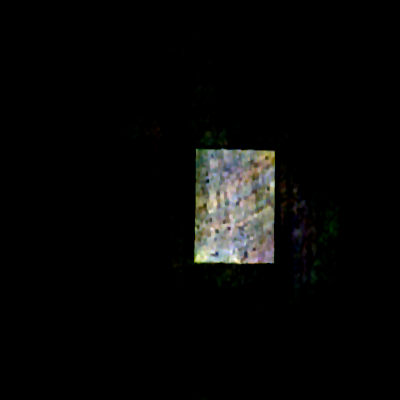} &
		\includegraphics[width = \resImgSz\columnwidth]{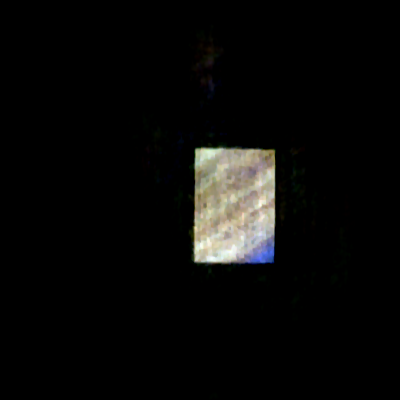} &
		\includegraphics[width = \resImgSz\columnwidth]{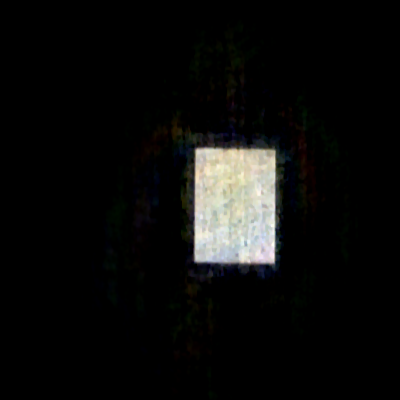} \\
        
        \rotatebox{90}{\parbox[c]{\rotateshift}{\centering simulated mask  }} &
        \includegraphics[width = \resImgSz\columnwidth]{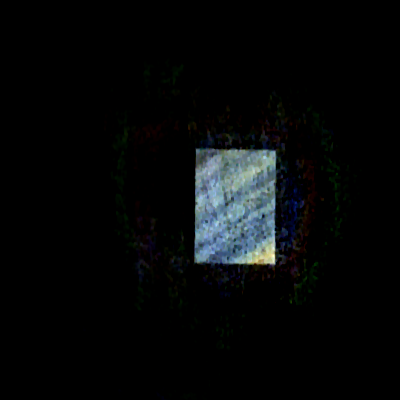} &
		\includegraphics[width = \resImgSz\columnwidth]{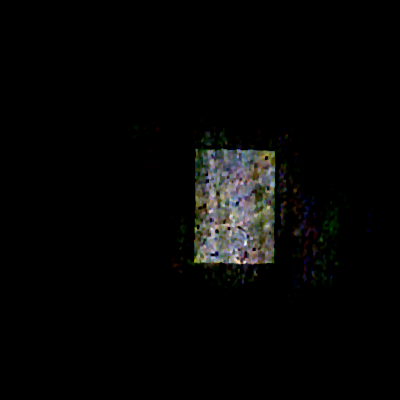} &
		\includegraphics[width = \resImgSz\columnwidth]{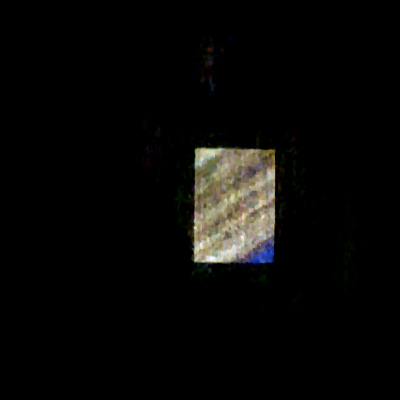} &
		\includegraphics[width = \resImgSz\columnwidth]{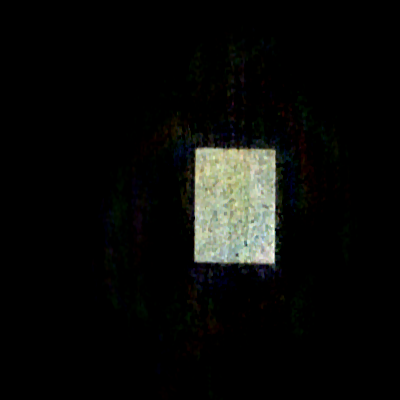} \\

	    \rotatebox{90}{\parbox[c]{\rotateshift}{\centering reference}} &
		\includegraphics[width = \resImgSz\columnwidth]{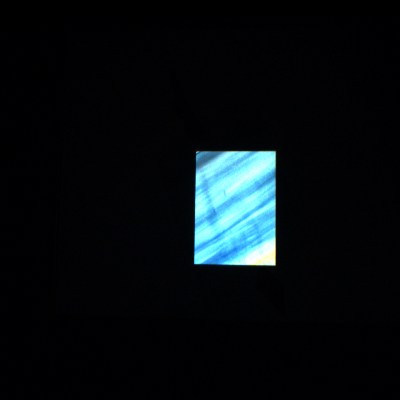} &
		\includegraphics[width = \resImgSz\columnwidth]{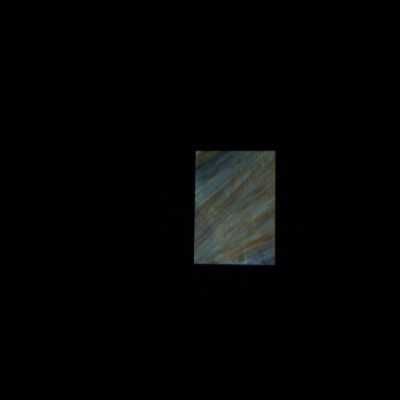} &
		\includegraphics[width = \resImgSz\columnwidth]{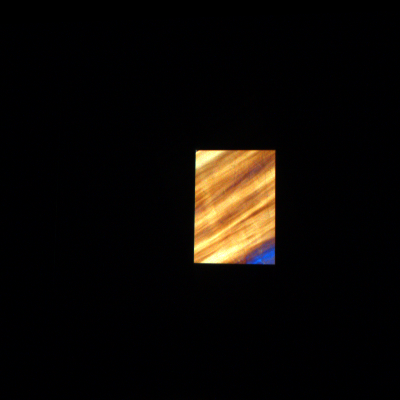} &
		\includegraphics[width = \resImgSz\columnwidth]{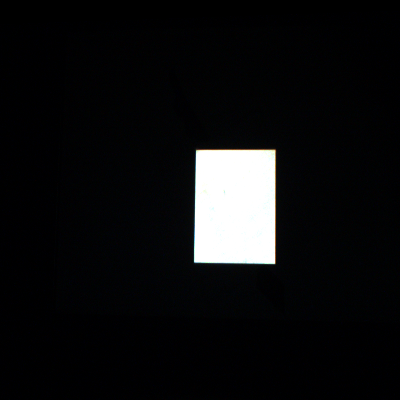} \\

	    &
		{\small{(a) $I_{0}$}} &
		{\small{(b) $I_{45}$}} &
		{\small{(c) $I_{90}$}} &
		{\small{(d) $I_{135}$}} \\

\end{tabular}
\caption{\textbf{Back illumination - matched-forward-model simulation:}(top) lensless reconstruction using a measured polarization mask (matched setting),(middle) lensless reconstruction using an idealized polarization mask (matched setting), and (bottom) corresponding spatially aligned reference images acquired using a lens-based camera with an external rotating polarizer.}
\label{fig:polBackIllum}
\end{figure}

\subsection{Matched Forward-Model Setting}
\label{subsec:matched}
We first evaluate the proposed reconstruction framework under matched forward-model conditions, in which the polarization mask used for data generation is identical to the mask used during reconstruction.
This setting establishes an upper bound on achievable reconstruction quality and serves as a reference for the subsequent experiments on model mismatch and real-world acquisitions.

To generate matched-model data, we first capture four single-polarization lensless measurements at polarization angles
$0^\circ, 45^\circ, 90^\circ$, and $135^\circ$ using the lensless camera with the physical diffuser, while placing an external rotating linear polarizer in front of the camera and without the physical polarization mask.
We then synthetically apply the polarization-mask modulation by combining the four measurements according to the modulation term of the forward model in \Cref{eq:forward_model} and a known polarization mask pattern.
Subsequent reconstruction (solving \Cref{eq:costFunc}) is performed using the same polarization mask and the measured diffuser PSF, resulting in a matched forward model for data simulation and reconstruction.

\begin{figure}[tb]
\def\resImgSz{0.2}
\newcommand{\colw}{\resImgSz\columnwidth}
\centering
\begin{tabular}{@{}c c c c@{}}
    \includegraphics[width=\resImgSz\columnwidth]{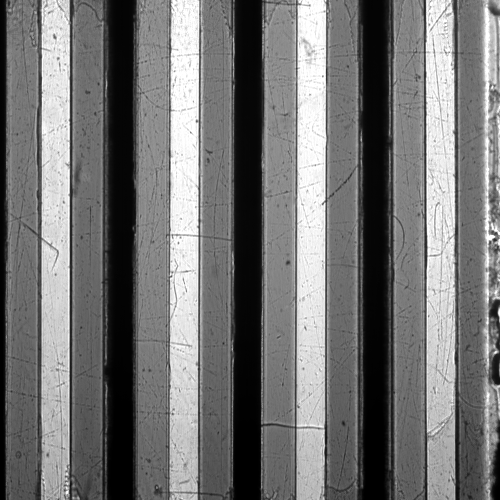} &
    \includegraphics[width=\resImgSz\columnwidth]{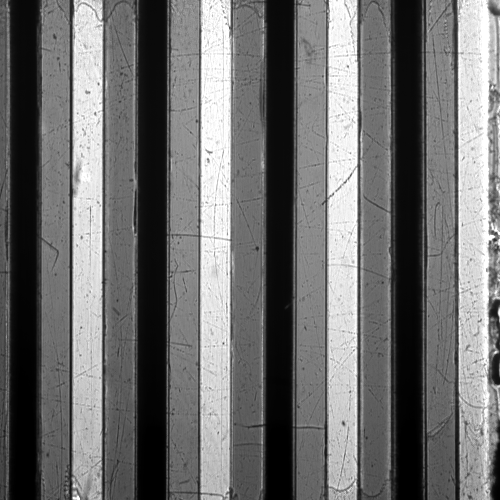} &
    \includegraphics[width=\resImgSz\columnwidth]{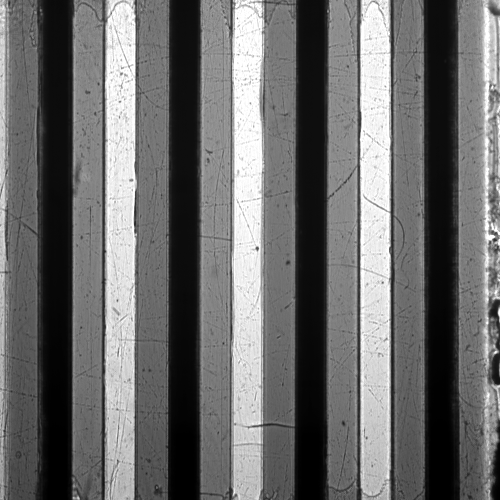} &
    \includegraphics[width=\resImgSz\columnwidth]{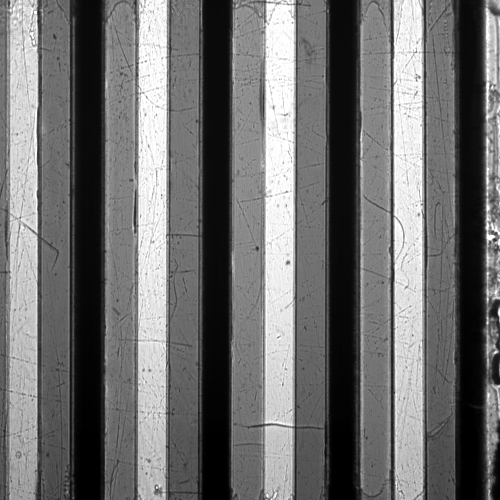} \\

    \makebox[\colw][c]{\small{(a) $I_{0}$}} &
    \makebox[\colw][c]{\small{(b) $I_{45}$}} &
    \makebox[\colw][c]{\small{(c) $I_{90}$}} &
    \makebox[\colw][c]{\small{(d) $I_{135}$}} \\

    \includegraphics[width=\resImgSz\columnwidth]{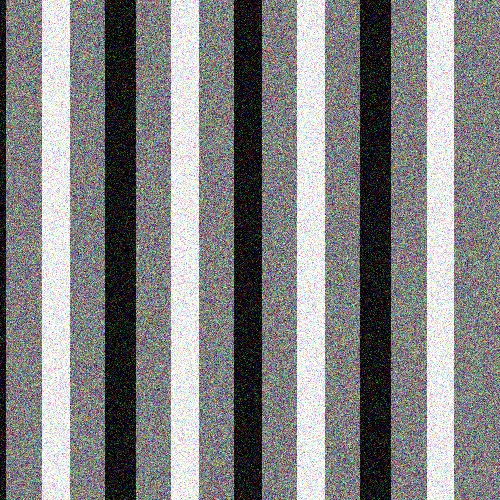} &
    \includegraphics[width=\resImgSz\columnwidth]{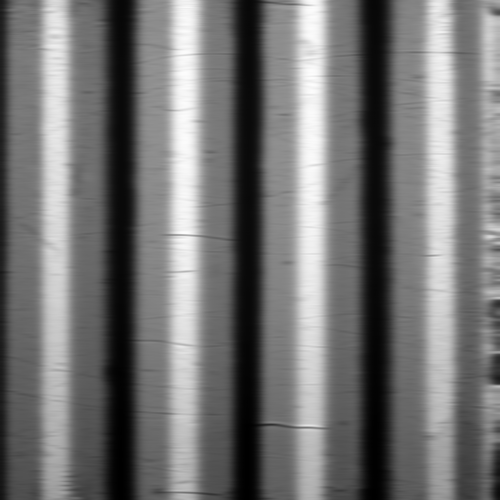} &
    \includegraphics[width=\resImgSz\columnwidth]{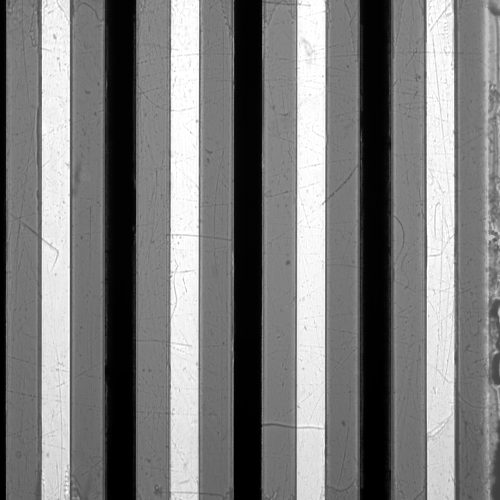} &
    \includegraphics[width=\resImgSz\columnwidth]{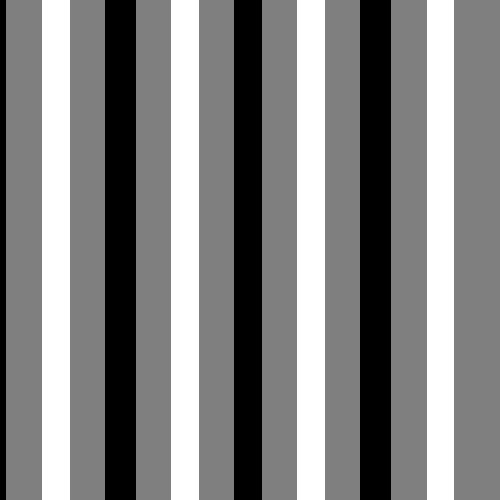} \\

    \makebox[\colw][c]{\small{(e) $I_{0}$ gaussian noise}} &
    \makebox[\colw][c]{\small{(f) $I_{0}$ gaussian blur}} &
    \makebox[\colw][c]{\small{(g) $I_{0}$ mean (meas. + sim.)}} &
    \makebox[\colw][c]{\small{(h) $I_{0}$  simulated mask}} \\
\end{tabular}

\caption{\textbf{Measured and perturbed polarization mask models used in the mismatch analysis.}
Top row: measured polarization mask responses corresponding to the four polarization orientations ($0^\circ,45^\circ,90^\circ,135^\circ$) obtained from the fabricated mask.
Bottom row: examples of perturbed mask models derived from the measured $0^\circ$ mask, including additive Gaussian noise, Gaussian blurring, interpolation with an idealized simulated mask (mean), and an idealized simulated mask.}
\label{fig:polmask}
\end{figure}

To examine the role of mask structure under matched settings, we consider two types of polarization masks: (i) a measured polarization mask obtained from the laboratory prototype (top row of \Cref{fig:polmask}), and (ii) an idealized mask generated in simulation.
An example of the simulated mask is shown for $I_{0}$ in \Cref{fig:polmask}(h). In both cases, the same mask is used consistently for data generation and reconstruction.

We evaluate the proposed polarization lensless imaging framework using both front-illuminated and back-illuminated scenes.
In the case of a front-illuminated scene, the scene is illuminated with linearly polarized light at $\theta=0^\circ$ from the left and similar linearly polarized light at $\theta=90^\circ$ from the right.
The results appear in \Cref{fig:sim-front-ill}, where source separation can be observed.
In this case, both data generation and reconstruction use the measured polarization mask.

The light-source separation under front illumination demonstrates the polarization imaging ability of the proposed scheme. In practice, polarization imaging is widely used with back-illumination for strain analysis of transparent materials, since mechanical stress induces birefringence and alters the polarization state of transmitted light.
As birefringence is often wavelength dependent, color–polarization imaging provides an effective means for visualizing and analyzing strain patterns.
We demonstrate this application using a back-illumination setting in which a plastic bag is placed in front of a linearly polarized LCD screen oriented at $\theta = 135^\circ$ (\Cref{fig:polBackIllum}). It can be seen that birefringence-induced effects are most pronounced in the $I_{0}$ and $I_{90}$ polarization channels, where structural details and color variations corresponding to strain direction and magnitude are well reconstructed. In contrast, the $I_{45}$ and $I_{135}$ channels exhibit weaker birefringence effects and are primarily dominated by lensless imaging artifacts.
In this example, results are shown for using either the measured physical mask or the idealized simulated mask.

Notably, strong reconstruction quality is observed for both the measured physical mask and the idealized simulated mask when each is used consistently for data generation and reconstruction.
This indicates that reconstruction fidelity under ideal conditions is not primarily determined by the perfection of the polarization mask itself, but rather by the accuracy of the assumed forward model. As long as the mask is accurately characterized and matched between data generation and reconstruction, the proposed framework supports high-quality snapshot polarization recovery.

\begin{figure}[t]
    \centering
    \includegraphics[width=0.98\columnwidth]{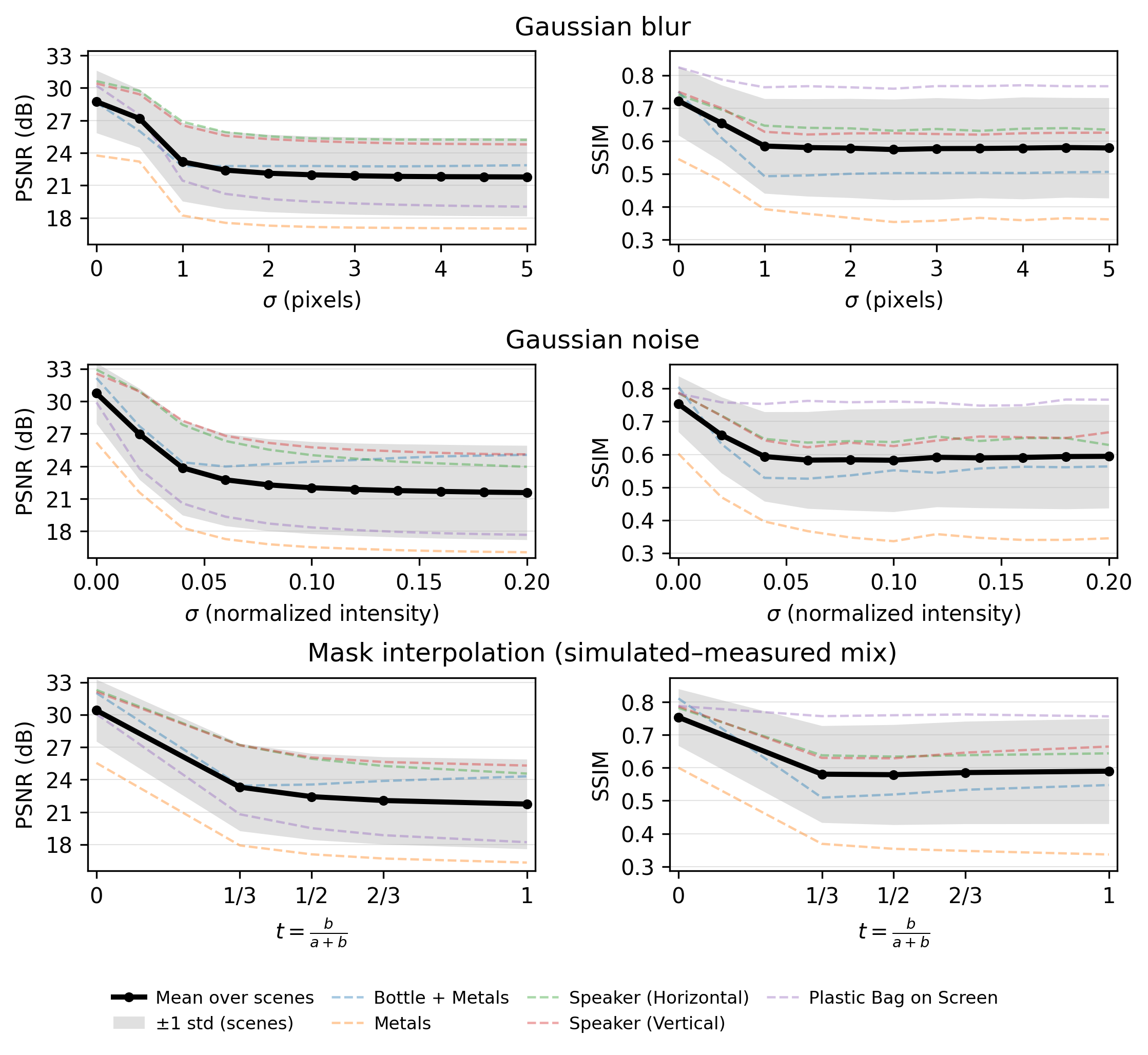}
    \caption{\textbf{Robustness of lensless polarization reconstruction to polarization-mask model mismatch.}
    Top: Gaussian blurring of the measured mask ($\sigma=0$ corresponds to the unblurred measured mask used for both data generation and reconstruction).
    Middle: additive Gaussian noise applied to the simulated mask ($\sigma=0$ corresponds to the noise-free simulated mask used for both data generation and reconstruction).
    Bottom: interpolation between an idealized simulated mask and the measured physical mask, with $t=\frac{b}{a+b}$ ($t=0$: normalized measured mask used for both data generation and reconstruction; $t=1$: simulated mask used for both).
    Left: PSNR. Right: SSIM.
    Solid curves show the mean across the scenes shown in \Cref{fig:sim-front-ill,fig:polBackIllum} with $\pm 1$ standard deviation shading; dashed curves denote individual scenes.}
\label{fig:robust-blur}
\vspace{-0.6em}
\end{figure}

Quantitatively, the matched-model setting (for both the measured and simulated masks) corresponds to the $\sigma=0$ and $t=0$ points in the robustness analysis (\Cref{fig:robust-blur}); the robustness experiment and the corresponding figure are described in \Cref{subsec:mismatch}.

\subsection{Model Mismatch Analysis}
\label{subsec:mismatch}
To characterize the sensitivity of lensless polarization reconstruction to polarization-mask mismatch in the forward model, we analyze reconstruction performance under controlled mask perturbations.

Data generation follows the same procedure as in \Cref{subsec:matched}. The key difference is that the polarization mask used to generate the measurements differs from the mask assumed during reconstruction. Throughout all mismatch experiments, all reconstructions use the measured diffuser PSF. This experimental design isolates the effect of polarization mask mismatch, which is the primary additional component in the proposed system.
Reconstruction quality is evaluated using PSNR and SSIM~\cite{wang2004image}, averaged over the four polarization sub-images.
Metrics are computed with respect to a lensless reference obtained by reconstructing the four polarization angles from measurements acquired without the polarization mask (using an external rotating polarizer), which isolates reconstruction error due to the polarization mask model-mismatch only. Regularization is held fixed across matched and mismatch experiments, and the observed degradation is therefore attributable to forward-model mismatch rather than changes in the reconstruction prior.

\noindent\textbf{Sources of mask mismatch in the lab prototype.}
In the lab prototype, the sensor cover glass enforces a gap between the sensor and the polarization mask. In addition, the glass substrate supporting the polarization stripes (located toward the diffuser) and the optical adhesive layer used during assembly further introduce deviations from the assumed forward model. Moreover, in-house fabrication introduces imperfections such as non-uniform stripe widths, non-ideal stripe edges, local stretching or deformation of the polarization film, surface scratches, residual adhesive between stripes, and small gaps or overlaps at stripe boundaries. Together, these effects lead to a substantial mismatch between the true mask modulation and the idealized modulation assumed by the forward model.

\noindent\textbf{Controlled mismatch modeling.}
Because of the imperfections described above, the measured polarization mask does not exactly follow the modulation assumed in the forward model. Since these effects cannot be fully simulated or incorporated into the forward model, we model mask mismatch using a small set of controlled and interpretable perturbations that approximate the dominant mismatch effects we assume in our laboratory prototype.

Specifically, we consider three representative forms of perturbations: spatially blurred masks (1D Gaussian blurring of the measured mask), noisy masks (additive Gaussian noise applied to the idealized simulated mask), and a convex interpolation between the idealized simulated mask and the measured mask (see \Cref{fig:polmask}).

Perturbations are applied to the mask in either the data-generation process or in the reconstruction model, such that the forward model assumed during reconstruction no longer matches the forward model used to generate the measurements. This setup mimics the uncertainty in the effective acquisition model available to the algorithm in a real imaging system.

For blur and interpolation, measurements are generated using a perturbed mask, while reconstruction is performed using the corresponding unperturbed mask (measured or idealized).
For noise, measurements are generated using the idealized (noise-free) mask, and additive noise is applied only to the reconstruction mask.

Blurring perturbation is introduced to approximate the spatial averaging and cross-talk caused by the finite distance between the polarization mask and the sensor pixel plane, as well as stripe-boundary effects in the laboratory prototype. In particular, after the modulation by the polarization mask, the diffuser-encoded light continues to propagate over an undesired small distance (due to the cover glass) before reaching the sensor. This diffraction and free-space propagation lead to additional spatial mixing between adjacent polarization stripes. Due to the striped geometry of the polarization mask, this propagation-induced mixing is expected to be predominantly anisotropic, with slightly stronger spatial averaging occurring perpendicular to the stripe direction. Accordingly, we model the blur as a one-dimensional Gaussian filter applied perpendicular to the polarization stripes. This effect is not inherent to the polarization multiplexing principle but arises from the non-zero distance between the mask and the sensor in the lab prototype. Although we do not explicitly model wave-optical propagation in the forward model, an analysis of the polarization mask as a diffraction grating shows that one-dimensional Gaussian blurring can be used as an interpretable approximation of these combined prototype-specific effects (See Supplement Document, \Cref{sec:appendix_mask_distance}).

The additive noise perturbation represents uncertainty in the estimated mask response, while the interpolation between simulated and measured masks captures systematic deviations arising from fabrication tolerances and non-ideal stripe modulation.

\Cref{fig:robust-blur} quantifies the effect of different forms of polarization mask mismatch on reconstruction quality.
Across all perturbation types, even modest deviations from the matched-mask setting lead to a pronounced drop in PSNR and SSIM, followed by a gradual saturation as mismatch severity increases. The observed saturation is influenced by the large dark background regions and by the fact that the foreground signal is already substantially degraded at high mismatch levels, reducing the sensitivity of PSNR and SSIM to further perturbations.
Importantly, spatial blur, additive noise, and mask-interpolation perturbations produce qualitatively similar degradation trends, suggesting that reconstruction quality is governed primarily by the degree of forward-model mismatch rather than its specific form.
While absolute performance varies across scenes, the overall trends are consistent, supporting the use of these controlled perturbations as interpretable proxies for a mismatch in the physical system.
Additional qualitative examples provided in the Supplement Document (\Cref{fig:mask-mismatch}) further show that mask mismatch induces artifacts similar to those observed in the lab prototype results.

Overall, the controlled perturbations considered in this analysis approximate dominant deviations of the real prototype from the assumed polarization-mask model and demonstrate that such deviations can substantially degrade reconstruction quality, providing a principled explanation of the real-world performance trends observed in  \Cref{subsec:real-world}.

\subsection{Real-World Experiments}
\label{subsec:real-world}
We evaluate real-world acquisitions obtained with the lab prototype to assess the practical performance of the proposed lensless polarization imaging system and to examine deviations from the assumed forward model.

\begin{figure}[t]
\centering
\setlength{\tabcolsep}{3pt}        %
\renewcommand{\arraystretch}{1.05} %

\begin{tabular}{>{\centering\arraybackslash}m{0.04\textwidth}
                *{4}{>{\centering\arraybackslash}m{0.21\columnwidth}}}

\multirow{2}{*}{\raisebox{-0.55\height}{\rotatebox{90}{\small\textbf{Real}}}} &
\includegraphics[width=\linewidth]{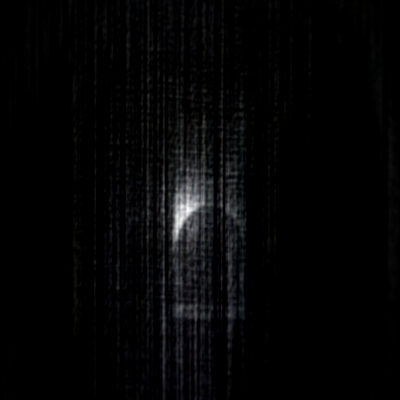} &
\includegraphics[width=\linewidth]{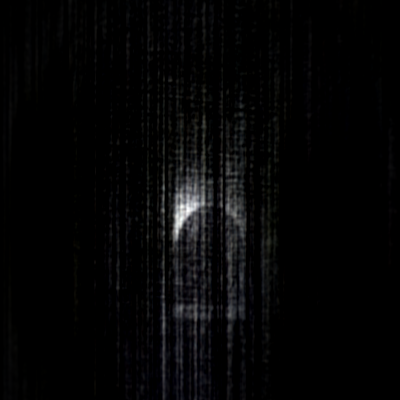} &
\includegraphics[width=\linewidth]{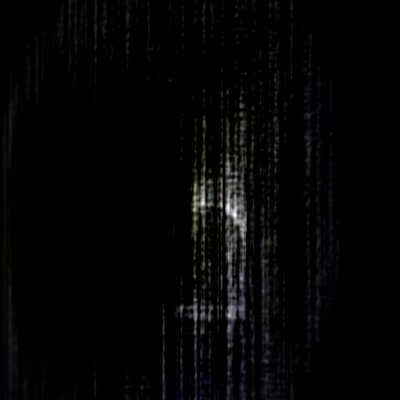} &
\includegraphics[width=\linewidth]{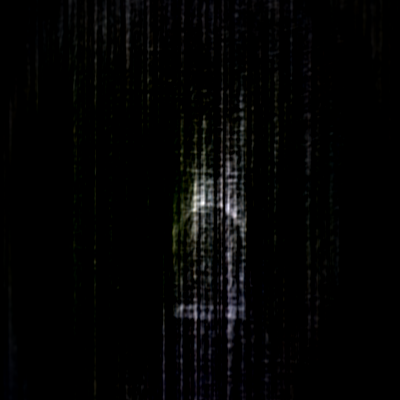} \\[2pt]

\multirow{2}{*}{\raisebox{-0.55\height}{\rotatebox{90}{\small\textbf{Sim}}}} 

& \includegraphics[width=\linewidth]{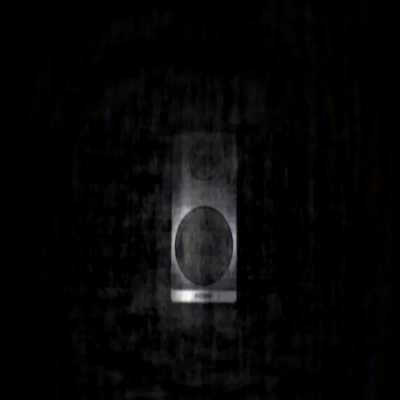} &
  \includegraphics[width=\linewidth]{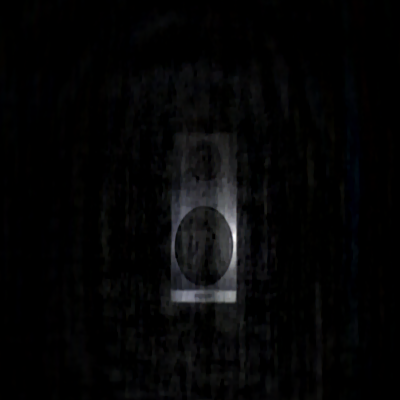} &
  \includegraphics[width=\linewidth]{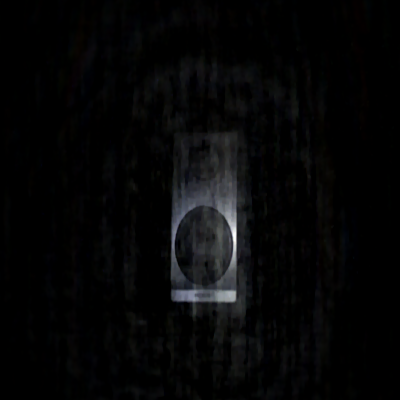} &
  \includegraphics[width=\linewidth]{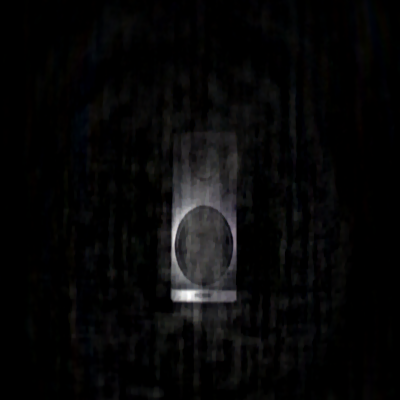} \\[-2pt]

& {\small $I_{0}$} & {\small $I_{45}$} & {\small $I_{90}$} & {\small $I_{135}$}

\end{tabular}
\caption{\textbf{Front illumination example.}
(top) Lensless polarization reconstruction from a single snapshot acquired with the lab prototype.
(bottom) Reconstruction of matched-mask setting obtained by synthetically multiplexing polarization lensless measurements using the measured polarization mask.
The real and simulated scenes are similar but not identical.
}
\label{fig:polFrontIllum}
\end{figure}

\Cref{fig:polFrontIllum} shows a representative front-illumination scene.
The top row presents polarization reconstructions obtained from a single snapshot acquired with the lab prototype, while the bottom row shows the corresponding reconstruction from synthetically multiplexed lensless measurements using the measured polarization mask, representing the matched-model reference.
Because the polarization mask had to be removed and reinstalled between acquisitions, the real and simulated scenes are similar but not identical, and the comparison is therefore qualitative.

This comparison highlights the gap between simulated and experimental performance, arising from the mismatch between the assumed and effective polarization mask forward model. Additional scenes are shown in the Supplement Document (\Cref{fig:sim-back-ill-appendix} and \Cref{fig:sim-front-ill-appendix}).

The overall scene structure and polarization-dependent illumination patterns are recovered in the lab prototype reconstructions, but with reduced contrast and increased artifacts compared to the ideal model-matched setting.
In particular, while the $I_{45}$ and $I_{135}$ channels exhibit approximately uniform illumination and the $I_{0}$ and $I_{90}$ channels show directional illumination effects, these polarization cues are attenuated in the lab acquisition.

This behavior is consistent with spatial mixing, polarization cross-talk, and other prototype-specific deviations introduced by the fabricated polarization mask, and is qualitatively consistent with the artifacts induced by the perturbations in the controlled mismatch experiments (see Supplement Document Fig.~S1), supporting polarization-mask mismatch as a major contributor to the observed sim--real performance gap.

These results represent the best reconstruction quality we obtained with the current prototype.
With improved fabrication and integration, such as on-chip or co-fabricated polarization masks that are precisely aligned with the sensor plane and minimize mask–sensor distance, the reconstruction quality is expected to approach the ideal performance demonstrated in simulation.

\section{Discussion and Limitations}
\label{sec:discussion}
The proposed lensless polarization camera demonstrates that snapshot recovery of RGB linear polarization images is feasible using a diffuser and a simple striped polarization multiplexing mask when the forward model is accurately characterized.

At the same time, the experimental results reveal a pronounced gap between matched-model simulations and real-world prototype performance. This gap does not stem from a fundamental limitation of the reconstruction algorithm or the multiplexing scheme, but rather from deviations between the assumed forward model and the effective physical acquisition process, due to lab manufacturing limitations and imperfections. The introduction of a polarization mask increases the structural complexity of the forward model, thereby amplifying sensitivity to modeling inaccuracies.

Our controlled mismatch analysis identifies polarization-mask mismatch as a dominant factor limiting system performance. Finite mask–sensor distance (e.g., due to the cover glass), fabrication imperfections, stripe boundary effects, and assembly tolerances introduce structured deviations from the idealized polarization modulation assumed during reconstruction, leading to spatial mixing and other distortions.
The close qualitative correspondence between artifacts observed in the physical prototype reconstructions and those induced by the controlled mask perturbations supports this conclusion.

Rather than proposing a new reconstruction strategy to compensate for these effects, this work characterizes the limits imposed by polarization mask forward-model mismatch. By decoupling optical feasibility from reconstruction robustness, the analysis provides a principled explanation for the observed simulation-to-real-world performance gap and indicates that increasing reconstruction complexity alone cannot compensate for systematic optical mismatch.

These findings motivate higher-precision polarization-mask fabrication and tighter integration with the sensor as a primary pathway for improving reconstruction fidelity. On-chip or co-fabricated implementations are expected to reduce system variability and bring the effective acquisition process closer to the assumed forward model, enabling performance to approach the ideal behavior observed in simulation.

Finally, we use ADMM with weighted anisotropic TV throughout, as it provides a stable and interpretable baseline under mismatch. While learning-based or hybrid priors may improve performance under perfectly matched conditions, our results indicate that such methods are particularly sensitive to forward-model inaccuracies, motivating the use of physics-based baselines when system variability is significant.

\section{Conclusion}
\label{sec:conclusion}
We proposed and analyzed a lensless polarization camera based on a diffuser and a striped polarization mask, and validated the system through both simulation and a laboratory prototype.
The random diffuser enabled image reconstruction from partially sampled measurements, which allowed polarization multiplexing directly in the image plane using a simple spatial division strategy.
The design enables a compact and lightweight camera, thus opening possibilities for polarization imaging in applications with extreme volume/weight requirements.

Experimental results demonstrate the potential of the proposed system for recovering polarization-dependent scene information, including polarization angle separation and birefringence-based strain analysis in transparent materials.
By extending the polarization mask structure to multiplex additional polarization states (e.g., circular polarization components), full-Stokes imaging can in principle be achieved.

While the performance of the current lab prototype is limited by practical mask–sensor integration and fabrication imperfections, our sensitivity analysis with respect to forward-model accuracy highlights a clear and actionable path toward improved performance. In particular, tighter integration of the polarization mask with the sensor, such as on-chip or co-fabricated implementations, offers a promising route toward reducing model mismatch and approaching the ideal performance observed under matched simulation conditions.

\newpage
\bibliographystyle{unsrtnat}
\bibliography{sample}

@article{polImagOld,
author = {Jerry E. Solomon},
journal = {Appl. Opt.},
number = {9},
pages = {1537--1544},
publisher = {OSA},
title = {Polarization imaging},
volume = {20},
month = {May},
year = {1981},
url = {http://www.osapublishing.org/ao/abstract.cfm?URI=ao-20-9-1537},
doi = {10.1364/AO.20.001537},
}

@article{kamilov2016parallel,
  title={A parallel proximal algorithm for anisotropic total variation minimization},
  author={Kamilov, Ulugbek S},
  journal={IEEE Transactions on Image Processing},
  volume={26},
  number={2},
  pages={539--548},
  year={2016},
  publisher={IEEE}
}

@article{baek2022lensless,
  title   = {Lensless polarization camera for single-shot full-{S}tokes imaging},
  author  = {Baek, Nakkyu and Lee, Yujin and Kim, Taeyoung and Jung, Jaewoo and Lee, Seung Ah},
  journal={APL Photonics},
  volume  = {7},
  number  = {11},
  year    = {2022}
}

@article{capassoPolCam,
author = {Noah A. Rubin  and Gabriele D’Aversa  and Paul Chevalier  and Zhujun Shi  and Wei Ting Chen  and Federico Capasso },
title = {Matrix Fourier optics enables a compact full-Stokes polarization camera},
journal = {Science},
volume = {365},
number = {6448},
pages = {eaax1839},
year = {2019},
doi = {10.1126/science.aax1839},

URL = {https://www.science.org/doi/abs/10.1126/science.aax1839},
eprint = {https://www.science.org/doi/pdf/10.1126/science.aax1839}
}

@article{zheng2025exploiting,
  title={Exploiting Scattering-Based Point Spread Functions for Snapshot 5D and Modality-Switchable Lensless Imaging},
  author={Zheng, Ze and Liu, Baolei and Song, Jiaqi and Zhu, Muchen and Han, Chi and Wang, Yao and Tian, Menghan and Xiong, Ying and Yang, Zhaohua and Zhong, Xiaolan and others},
  journal={Laser \& Photonics Reviews},
  pages={e01888},
  year={2025},
  publisher={Wiley Online Library}
}

@inproceedings{zheng2023spectral,
  title     = {Spectral and polarization imaging by a lensless diffuser camera},
  author    = {Zheng, Ze and Liu, Baolei and Tian, Menghan and Wang, Yao and Song, Jiaqi and Shan, Xuchen and Zhong, Xiaolan and Wang, Fan},
  publisher = {{SPIE}},
  volume    = {12975},
 booktitle={6th Optics Young Scientist Summit (OYSS 2023)},
  pages     = {217--220},
  year      = {2023}
}

@article{polDenoise_PCA,
author = {Junchao Zhang and Haibo Luo and Rongguang Liang and Wei Zhou and Bin Hui and Zheng Chang},
journal = {Opt. Express},
number = {3},
pages = {2391--2400},
publisher = {OSA},
title = {PCA-based denoising method for division of focal plane polarimeters},
volume = {25},
month = {Feb},
year = {2017},
url = {http://www.osapublishing.org/oe/abstract.cfm?URI=oe-25-3-2391},
doi = {10.1364/OE.25.002391},
}

@article{polMicroscopy,
author = {Shribak, Michael},
year = {2015},
month = {11},
pages = {17340},
title = {Polychromatic polarization microscope: bringing colors to a colorless world},
volume = {5},
journal = {Scientific reports},
doi = {10.1038/srep17340}
}

@article{polDenoiseLearn,
author = {Xiaobo Li and Haiyu Li and Yang Lin and Jianhua Guo and Jingyu Yang and Huanjing Yue and Kun Li and Chuan Li and Zhenzhou Cheng and Haofeng Hu and Tiegen Liu},
journal = {Opt. Express},
number = {11},
pages = {16309--16321},
publisher = {OSA},
title = {Learning-based denoising for polarimetric images},
volume = {28},
month = {May},
year = {2020},
url = {http://www.osapublishing.org/oe/abstract.cfm?URI=oe-28-11-16309},
doi = {10.1364/OE.391017},
}

@article{stokesParamPaper,
author = {Perrin,Francis },
title = {Polarization of Light Scattered by Isotropic Opalescent Media},
journal = {The Journal of Chemical Physics},
volume = {10},
number = {7},
pages = {415-427},
year = {1942},
doi = {10.1063/1.1723743},

URL = { 
        https://doi.org/10.1063/1.1723743
    
},
eprint = { 
        https://doi.org/10.1063/1.1723743
    
}

}

@INPROCEEDINGS{zimpol,
       author = {{Schmid}, H.~M. and {Gisler}, D. and {Joos}, F. and {Povel}, H.~P. and {Stenflo}, J.~O. and {Feldt}, M. and {Lenzen}, R. and {Brandner}, W. and {Tinbergen}, J. and {Quirrenbach}, A. and {Stuik}, R. and {Gratton}, R. and {Turatto}, M. and {Neuh{\"a}user}, R.},
        title = "{ZIMPOL/CHEOPS: a Polarimetric Imager for the Direct Detection of Extra-solar Planets}",
    booktitle = {Astronomical Polarimetry: Current Status and Future Directions},
         year = 2005,
       editor = {{Adamson}, A. and {Aspin}, C. and {Davis}, C. and {Fujiyoshi}, T.},
       series = {Astronomical Society of the Pacific Conference Series},
       volume = {343},
        month = dec,
        pages = {89},
       adsurl = {https://ui.adsabs.harvard.edu/abs/2005ASPC..343...89S}
}

@article{polImg_medical,
	doi = {10.1088/2040-8986/abbf8a},
	url = {https://doi.org/10.1088/2040-8986/abbf8a},
	year = 2020,
	month = {nov},
	publisher = {{IOP} Publishing},
	volume = {22},
	number = {12},
	pages = {123001},
	author = {Jessica C Ramella-Roman and Ilyas Saytashev and Mattia Piccini},
	title = {A review of polarization-based imaging technologies for clinical and preclinical applications},
	journal = {Journal of Optics}
}

@article{polDem_CNN,
author = {Junchao Zhang and Jianbo Shao and Haibo Luo and Xiangyue Zhang and Bin Hui and Zheng Chang and Rongguang Liang},
journal = {Opt. Lett.},
number = {18},
pages = {4534--4537},
publisher = {OSA},
title = {Learning a convolutional demosaicing network for microgrid polarimeter imagery},
volume = {43},
month = {Sep},
year = {2018},
url = {http://www.osapublishing.org/ol/abstract.cfm?URI=ol-43-18-4534},
doi = {10.1364/OL.43.004534},
}

@ARTICLE{flatCam,
  author={M. S. {Asif} and A. {Ayremlou} and A. {Sankaranarayanan} and A. {Veeraraghavan} and R. G. {Baraniuk}},
  journal={IEEE Transactions on Computational Imaging}, 
  title={FlatCam: Thin, Lensless Cameras Using Coded Aperture and Computation}, 
  year={2017},
  volume={3},
  number={3},
  pages={384-397},}

@article{DiffuserCam,
author = {Nick Antipa and Grace Kuo and Reinhard Heckel and Ben Mildenhall and Emrah Bostan and Ren Ng and Laura Waller},
journal = {Optica},
number = {1},
pages = {1--9},
publisher = {OSA},
title = {DiffuserCam: lensless single-exposure 3D imaging},
volume = {5},
month = {Jan},
year = {2018},
url = {http://www.osapublishing.org/optica/abstract.cfm?URI=optica-5-1-1},
doi = {10.1364/OPTICA.5.000001},
}

@ARTICLE{PhlatCam,
  author={V. {Boominathan} and J. {Adams} and J. {Robinson} and A. {Veeraraghavan}},
  journal={IEEE Transactions on Pattern Analysis and Machine Intelligence}, 
  title={PhlatCam: Designed phase-mask based thin lensless camera}, 
  year={2020},
  volume={},
  number={},
  pages={1-1},}

@INPROCEEDINGS{WallerRlngShtr,
  author={N. {Antipa} and P. {Oare} and E. {Bostan} and R. {Ng} and L. {Waller}},
  booktitle={2019 IEEE International Conference on Computational Photography (ICCP)}, 
  title={Video from Stills: Lensless Imaging with Rolling Shutter}, 
  year={2019},
  volume={},
  number={},
  pages={1-8},}

@article{diffSpec,
author = {Kristina Monakhova and Kyrollos Yanny and Neerja Aggarwal and Laura Waller},
journal = {Optica},
number = {10},
pages = {1298--1307},
publisher = {OSA},
title = {Spectral DiffuserCam: lensless snapshot hyperspectral imaging with a spectral filter array},
volume = {7},
month = {Oct},
year = {2020},
url = {http://www.osapublishing.org/optica/abstract.cfm?URI=optica-7-10-1298},
doi = {10.1364/OPTICA.397214},
}

@inproceedings{elmalem2021lensless,
  title={A lensless polarization camera},
  author={Elmalem, Shay and Giryes, Raja},
  booktitle={Computational Optical Sensing and Imaging},
  pages={CTh7A--1},
  year={2021},
  organization={Optica Publishing Group}
}

@article{liu2021high,
  title={High-efficient and high-accurate integrated division-of-time polarimeter},
  author={Liu, Wei and Liao, Jiawen and Yu, Yu and Zhang, Xinliang},
  journal={APL Photonics},
  volume={6},
  number={7},
  year={2021},
  publisher={AIP Publishing}
}

@inproceedings{salsaPolCam,
author = {Nicolas Lefaudeux and Nicolas Lechocinski and Sebastien Breugnot and Philippe Clemenceau},
title = {{Compact and robust linear Stokes polarization camera}},
volume = {6972},
booktitle = {Polarization: Measurement, Analysis, and Remote Sensing VIII},
editor = {David B. Chenault and Dennis H. Goldstein},
organization = {International Society for Optics and Photonics},
publisher = {{SPIE}},
pages = {76 -- 87},
year = {2008},
doi = {10.1117/12.781876},
URL = {https://doi.org/10.1117/12.781876}
}

@misc{lucidPolCam,
  author       = {Lucid Vision Labs},
  title        = {Polarization Cameras Based on On-Chip Polarization Sensors},
  howpublished = {\url{https://thinklucid.com/polarized-camera-resource-center/}},
  note         = {Accessed: 2026-01}
}

@misc{sonyPol,
  author       = {{Sony Semiconductor Solutions Corporation}},
  title        = {Polarization Image Sensor Technology (Polarsens)},
  howpublished = {\url{https://www.sony-semicon.com/en/technology/industry/polarsens.html}},
  note         = {Accessed: 2026-01},
  year         = {2026}
}

@article{liu2024review,
  title={Review of polarimetric image denoising},
  author={Liu, Hedong and Li, Xiaobo and Wang, Zihan and Huang, Yizhao and Zhai, Jingsheng and Hu, Haofeng and others},
  journal={Advanced Imaging},
  volume={1},
  number={2},
  pages={022001},
  year={2024},
  publisher={Editorial Office of Advanced Imaging}
}

@article{zheng2024color,
  title={Color polarization imaging demosaicing based on Stokes vector information complementation and fusion},
  author={Zheng, Yubo and Zhang, Xiangyue and Wu, Chengdong and Ji, Peng and Ru, Jingyu},
  journal={Optics Express},
  volume={32},
  number={25},
  pages={44049--44066},
  year={2024},
  publisher={Optica Publishing Group}
}

@article{huang2023high,
  title={High-resolution metalens imaging polarimetry},
  author={Huang, Zhaorui and Zheng, Yaqin and Li, Junhao and Cheng, Yongzhi and Wang, Jian and Zhou, Zhang-Kai and Chen, Lin},
  journal={Nano letters},
  volume={23},
  number={23},
  pages={10991--10997},
  year={2023},
  publisher={ACS Publications}
}

@article{sasagawa2022polarization,
  title={Polarization image sensor for highly sensitive polarization modulation imaging based on stacked polarizers},
  author={Sasagawa, Kiyotaka and Okada, Ryoma and Haruta, Makito and Takehara, Hironari and Tashiro, Hiroyuki and Ohta, Jun},
  journal={IEEE Transactions on Electron Devices},
  volume={69},
  number={6},
  pages={2924--2931},
  year={2022},
  publisher={IEEE}
}

@article{li2025flat,
  title={Flat, wide field-of-view imaging polarimeter},
  author={Li, Lisa W and Oh, Jaewon and Miller, Harris and Capasso, Federico and Rubin, Noah A},
  journal={Optica},
  volume={12},
  number={6},
  pages={799--811},
  year={2025},
  publisher={Optica Publishing Group}
}

@inproceedings{zhou2025pidsr,
  title={PIDSR: Complementary Polarized Image Demosaicing and Super-Resolution},
  author={Zhou, Shuangfan and Zhou, Chu and Lyu, Youwei and Guo, Heng and Ma, Zhanyu and Shi, Boxin and Sato, Imari},
  booktitle={Proceedings of the Computer Vision and Pattern Recognition Conference},
  pages={16081--16090},
  year={2025}
}

@inproceedings {KaustPolDem,
booktitle = {Vision, Modeling and Visualization},
editor = {Schulz, Hans-Jörg and Teschner, Matthias and Wimmer, Michael},
title = {{Polarization Demosaicking for Monochrome and Color Polarization Focal Plane Arrays}},
author = {Qiu, Simeng and Fu, Qiang and Wang, Congli and Heidrich, Wolfgang },
year = {2019},
publisher = {The Eurographics Association},
ISBN = {978-3-03868-098-7},
DOI = {10.2312/vmv.20191325}
}

@article{lenslessMenon,
author = {Ganghun Kim and Kyle Isaacson and Rachael Palmer and Rajesh Menon},
journal = {Appl. Opt.},
number = {23},
pages = {6450--6456},
publisher = {OSA},
title = {Lensless photography with only an image sensor},
volume = {56},
month = {Aug},
year = {2017},
url = {http://www.osapublishing.org/ao/abstract.cfm?URI=ao-56-23-6450},
doi = {10.1364/AO.56.006450},
}

@article{lenslessBarbasthatis,
author = {Ayan Sinha and Justin Lee and Shuai Li and George Barbastathis},
journal = {Optica},
number = {9},
pages = {1117--1125},
publisher = {OSA},
title = {Lensless computational imaging through deep learning},
volume = {4},
month = {Sep},
year = {2017},
url = {http://www.osapublishing.org/optica/abstract.cfm?URI=optica-4-9-1117},
doi = {10.1364/OPTICA.4.001117},
}

@article{poincare1892theorie,
  title={Th{\'e}orie Math{\'e}matique de la Lumi{\`e}re, Vol. 2 (Georges Carr{\'e}, Paris)},
  author={Poincar{\'e}, H},
  journal={MI MISHCHENKO AND LD TRAVIS},
  volume={44},
  year={1892}
}

@article{WallerMiniScope,
title = {Miniscope3D: optimized single-shot miniature 3D fluorescence microscopy},
author = {Kyrollos Yanny and Nick Antipa and William Liberti and Sam Dehaeck and Kristina Monakhova and Fanglin Linda Liu and Konlin Shen and Ren Ng and Laura Waller},
url = {https://www.nature.com/articles/s41377-020-00403-7},
doi = {https://doi.org/10.1038/s41377-020-00403-7},
year = {2020},
date = {2020-10-02},
journal = {Light: Science \& Applications},
volume = {9},
number = {171},
pubstate = {published},
tppubtype = {article}
}

@article{boominathan2021recent,
  title={Recent advances in lensless imaging},
  author={Boominathan, Vivek and Robinson, Jacob T and Waller, Laura and Veeraraghavan, Ashok},
  journal={Optica},
  volume={9},
  number={1},
  pages={1--16},
  year={2021},
  publisher={Optical Society of America}
}

@article{monakhova2019learned,
  title={Learned reconstructions for practical mask-based lensless imaging},
  author={Monakhova, Kristina and Yurtsever, Joshua and Kuo, Grace and Antipa, Nick and Yanny, Kyrollos and Waller, Laura},
  journal={Optics express},
  volume={27},
  number={20},
  pages={28075--28090},
  year={2019},
  publisher={Optical Society of America}
}

@article{bezzam2025towards,
  title={Towards Robust and Generalizable Lensless Imaging With Modular Learned Reconstruction},
  author={Bezzam, Eric and Perron, Yohann and Vetterli, Martin},
  journal={IEEE Transactions on Computational Imaging},
  volume={11},
  pages={213--227},
  year={2025},
  publisher={IEEE}
}

@article{khan2020flatnet,
  title={Flatnet: Towards photorealistic scene reconstruction from lensless measurements},
  author={Khan, Salman Siddique and Sundar, Varun and Boominathan, Vivek and Veeraraghavan, Ashok and Mitra, Kaushik},
  journal={IEEE Transactions on Pattern Analysis and Machine Intelligence},
  volume={44},
  number={4},
  pages={1934--1948},
  year={2020},
  publisher={IEEE}
}

@inproceedings{hung2025scalable,
  title     = {Scalable dataset acquisition for data-driven lensless imaging},
  author    = {Hung, Clara S and Kabuli, Leyla A and Ponomarenko, Vasilisa and Waller, Laura},
  booktitle = {Computational Optical Imaging and Artificial Intelligence in Biomedical Sciences II},
  volume    = {13333},
  pages     = {54--58},
  year      = {2025},
  publisher = {{SPIE}}
}

@article{yosef2024difuzcam,
  title={Difuzcam: Replacing camera lens with a mask and a diffusion model},
  author={Yosef, Erez and Giryes, Raja},
  journal={arXiv preprint arXiv:2408.07541},
  year={2024}
}

@article{guan2025review,
  title={Review of polarization-based technology for biomedical applications},
  author={Guan, Caizhong and Zeng, Nan and He, Honghui},
  journal={Journal of Innovative Optical Health Sciences},
  volume={18},
  number={02},
  pages={2430002},
  year={2025},
  publisher={World Scientific}
}

@article{li2025robust,
  title={A robust underwater polarization image recovery based on Angle of Polarization with low-rank and sparse decomposition},
  author={Li, Yafeng and Chen, Yuehan and Zhang, Jiqing and Li, Yudong and Tang, Haoming and Fu, Xianping},
  journal={Optics \& Laser Technology},
  volume={181},
  pages={111669},
  year={2025},
  publisher={Elsevier}
}

@article{underwaterPolImg,
author = {Yanmin Zhu and Tianjiao Zeng and Kewei Liu and Zhenbo Ren and Edmund Y. Lam},
journal = {Opt. Express},
number = {25},
pages = {41865--41881},
publisher = {OSA},
title = {Full scene underwater imaging with polarization and an untrained network},
volume = {29},
month = {Dec},
year = {2021},
url = {http://www.osapublishing.org/oe/abstract.cfm?URI=oe-29-25-41865},
doi = {10.1364/OE.444755},
}

@article{fogPolImg,
author = {Julien Fade and Swapnesh Panigrahi and Anthony Carr\'{e} and Ludovic Frein and Cyril Hamel and Fabien Bretenaker and Hema Ramachandran and Mehdi Alouini},
journal = {Appl. Opt.},
number = {18},
pages = {3854--3865},
publisher = {OSA},
title = {Long-range polarimetric imaging through fog},
volume = {53},
month = {Jun},
year = {2014},
url = {http://www.osapublishing.org/ao/abstract.cfm?URI=ao-53-18-3854},
doi = {10.1364/AO.53.003854},
}

@InProceedings{deepSfp,
author="Ba, Yunhao
and Gilbert, Alex
and Wang, Franklin
and Yang, Jinfa
and Chen, Rui
and Wang, Yiqin
and Yan, Lei
and Shi, Boxin
and Kadambi, Achuta",
editor="Vedaldi, Andrea
and Bischof, Horst
and Brox, Thomas
and Frahm, Jan-Michael",
title="Deep Shape from Polarization",
booktitle="Computer Vision -- ECCV 2020",
year="2020",
publisher="Springer International Publishing",
address="Cham",
pages="554--571",
isbn="978-3-030-58586-0"
}

@InProceedings{3dPolImg,
author = {Deschaintre, Valentin and Lin, Yiming and Ghosh, Abhijeet},
title = {Deep polarization imaging for 3D shape and SVBRDF acquisition},
booktitle = {Proceedings of the IEEE/CVF Conference on Computer Vision and Pattern Recognition (CVPR)},
month = {June},
year = {2021}
}

@article{remoteSensingPolImg,
author = {J. Scott Tyo and Dennis H. Goldstein and David B. Chenault and Joseph A. Shaw},
journal = {Appl. Opt.},
number = {22},
pages = {5451--5452},
publisher = {OSA},
title = {Polarization in Remote Sensing--introduction},
volume = {45},
month = {Aug},
year = {2006},
url = {http://www.osapublishing.org/ao/abstract.cfm?URI=ao-45-22-5451},
doi = {10.1364/AO.45.005451},
}

@article{tyo2006review,
  title={Review of passive imaging polarimetry for remote sensing applications},
  author={Tyo, J Scott and Goldstein, Dennis L and Chenault, David B and Shaw, Joseph A},
  journal={Applied optics},
  volume={45},
  number={22},
  pages={5453--5469},
  year={2006},
  publisher={Optical Society of America}
}

@book{goldstein2017polarized,
  title={Polarized light},
  author={Goldstein, Dennis H},
  year={2017},
  publisher={CRC press}
}

@article{wang2004image,
  title={Image quality assessment: from error visibility to structural similarity},
  author={Wang, Zhou and Bovik, Alan C and Sheikh, Hamid R and Simoncelli, Eero P},
  journal={IEEE transactions on image processing},
  volume={13},
  number={4},
  pages={600--612},
  year={2004},
  publisher={IEEE}
}

@InProceedings{DIP,
author = {Ulyanov, Dmitry and Vedaldi, Andrea and Lempitsky, Victor},
title = {Deep Image Prior},
booktitle = {Proceedings of the IEEE Conference on Computer Vision and Pattern Recognition (CVPR)},
month = {June},
year = {2018}
}

@inproceedings{raniwala2023improved,
  title={Improved fabrication and calibration for snapshot computational hyperspectral imaging},
  author={Raniwala, Y and Aggarwal, N and Waller, L},
  booktitle={Multiscale Imaging and Spectroscopy IV},
  volume={12363},
  pages={25--31},
  year={2023},
  publisher = {{SPIE}}
}

@article{wallerUntrained,
author = {Kristina Monakhova and Vi Tran and Grace Kuo and Laura Waller},
journal = {Opt. Express},
number = {13},
pages = {20913--20929},
publisher = {OSA},
title = {Untrained networks for compressive lensless photography},
volume = {29},
month = {Jun},
year = {2021},
url = {http://www.osapublishing.org/oe/abstract.cfm?URI=oe-29-13-20913},
doi = {10.1364/OE.424075},
}

@article{pistellato2022deep,
  title={Deep demosaicing for polarimetric filter array cameras},
  author={Pistellato, Mara and Bergamasco, Filippo and Fatima, Tehreem and Torsello, Andrea},
  journal={IEEE Transactions on Image Processing},
  volume={31},
  pages={2017--2026},
  year={2022},
  publisher={IEEE}
}

@article{zhou2025learning,
  title={Learning to deblur polarized images},
  author={Zhou, Chu and Teng, Minggui and Zhou, Xinyu and Xu, Chao and Sato, Imari and Shi, Boxin},
  journal={International Journal of Computer Vision},
  pages={1--16},
  year={2025},
  publisher={Springer}
}

@article{qian2024robust,
  title={Robust unrolled network for lensless imaging with enhanced resistance to model mismatch and noise},
  author={Qian, Hui and Ling, Hong and Lu, XiaoQiang},
  journal={Optics Express},
  volume={32},
  number={17},
  pages={30267--30283},
  year={2024},
  publisher={Optica Publishing Group}
}

@misc{ongie2020deep,
      title={Deep Learning Techniques for Inverse Problems in Imaging}, 
      author={Gregory Ongie and Ajil Jalal and Christopher A. Metzler and Richard G. Baraniuk and Alexandros G. Dimakis and Rebecca Willett},
      year={2020},
      eprint={2005.06001},
      archivePrefix={arXiv},
      primaryClass={eess.IV}
}

@article{RUDIN92Nonlinear,
title = {Nonlinear total variation based noise removal algorithms},
journal = {Physica D: Nonlinear Phenomena},
volume = {60},
number = {1},
pages = {259-268},
year = {1992},
author = {Leonid I. Rudin and Stanley Osher and Emad Fatemi},
}

@article{TokyoTech_polDem,
  title={Monochrome And Color Polarization Demosaicking Using Edge-Aware Residual Interpolation},
  author={Miki Morimatsu and Yusuke Monno and Masayuki Tanaka and M. Okutomi},
  journal={2020 IEEE International Conference on Image Processing (ICIP)},
  year={2020},
  pages={2571-2575}
}

@article{ADMM,
author = {Boyd, Stephen and Parikh, Neal and Chu, Eric and Peleato, Borja and Eckstein, Jonathan},
title = {Distributed Optimization and Statistical Learning via the Alternating Direction Method of Multipliers},
year = {2011},
issue_date = {January 2011},
publisher = {Now Publishers Inc.},
address = {Hanover, MA, USA},
volume = {3},
number = {1},
issn = {1935-8237},
url = {https://doi.org/10.1561/2200000016},
doi = {10.1561/2200000016},
journal = {Found. Trends Mach. Learn.},
month = {jan},
pages = {1–122},
numpages = {122}
}

@article{ polTissueRec,
author = {Carla Rodr\'{i}guez and Albert Van Eeckhout and Laia Ferrer and Enrique Garcia-Caurel and Emilio Gonz\'{a}lez-Arnay and Juan Campos and Angel Lizana},
journal = {Biomed. Opt. Express},
number = {8},
pages = {4852--4872},
publisher = {OSA},
title = {Polarimetric data-based model for tissue recognition},
volume = {12},
month = {Aug},
year = {2021},
url = {http://www.osapublishing.org/boe/abstract.cfm?URI=boe-12-8-4852},
doi = {10.1364/BOE.426387},
}

@article{Katz_lenslessEndoscopy_18,
author = {Uri Weiss and Ori Katz},
journal = {Opt. Express},
number = {22},
pages = {28808--28817},
publisher = {OSA},
title = {Two-photon lensless micro-endoscopy with in-situ wavefront correction},
volume = {26},
month = {Oct},
year = {2018},
url = {http://www.osapublishing.org/oe/abstract.cfm?URI=oe-26-22-28808},
doi = {10.1364/OE.26.028808},
}

@article{cai2024phocolens,
  title={Phocolens: Photorealistic and consistent reconstruction in lensless imaging},
  author={Cai, Xin and You, Zhiyuan and Zhang, Hailong and Gu, Jinwei and Liu, Wentao and Xue, Tianfan},
  journal={Advances in Neural Information Processing Systems},
  volume={37},
  pages={12219--12242},
  year={2024}
}

@article{zeng2021robust,
  title={Robust reconstruction with deep learning to handle model mismatch in lensless imaging},
  author={Zeng, Tianjiao and Lam, Edmund Y},
  journal={IEEE Transactions on Computational Imaging},
  volume={7},
  pages={1080--1092},
  year={2021},
  publisher={IEEE}
}

@article{burger2019convergence,
  title={Convergence rates and structure of solutions of inverse problems with imperfect forward models},
  author={Burger, Martin and Korolev, Yury and Rasch, Julian},
  journal={Inverse Problems},
  volume={35},
  number={2},
  pages={024006},
  year={2019},
  publisher={IOP Publishing}
}

@article{antun2020instabilities,
  title={On instabilities of deep learning in image reconstruction and the potential costs of AI},
  author={Antun, Vegard and Renna, Francesco and Poon, Clarice and Adcock, Ben and Hansen, Anders C},
  journal={Proceedings of the National Academy of Sciences},
  volume={117},
  number={48},
  pages={30088--30095},
  year={2020},
  publisher={National Academy of Sciences}
}

@book{bertero2021introduction,
  title={Introduction to inverse problems in imaging},
  author={Bertero, Mario and Boccacci, Patrizia and De Mol, Christine},
  year={2021},
  publisher={CRC press}
}

@article{Katz_lenslessFiberBundle,
author = {Amir Porat and Esben Ravn Andresen and Herv\'{e} Rigneault and Dan Oron and Sylvain Gigan and Ori Katz},
journal = {Opt. Express},
number = {15},
pages = {16835--16855},
publisher = {OSA},
title = {Widefield lensless imaging through a fiber bundle via speckle correlations},
volume = {24},
month = {Jul},
year = {2016},
url = {http://www.osapublishing.org/oe/abstract.cfm?URI=oe-24-15-16835},
doi = {10.1364/OE.24.016835},
}

\newpage

\appendix
\section*{Supplement Document}
\noindent This supplement document provides additional implementation details and qualitative results supporting the main paper.
We describe the physics-based ADMM reconstruction used throughout all experiments, including the weighted anisotropic total-variation prior (\Cref{sec:appendix_physics_based}).

Additional qualitative results further analyze the impact of polarization-mask mismatch (\Cref{sec:appendix_mismatch}) and examine the effect of the finite distance between the polarization mask and the sensor (\Cref{sec:appendix_mask_distance}).
We also present lensless polarization reconstructions from real-world acquisitions across multiple scenes and illumination conditions (\Cref{sec:appendix_additional_real}).

\section{ADMM Reconstruction Details}
\label{sec:appendix_physics_based}
This section provides additional implementation details for the physics-based ADMM reconstruction described in \Cref{subsec:reconstruction} of the main paper, which solves the polarization reconstruction problem defined in \Cref{eq:costFunc} using the forward operator in \Cref{eq:forward_model}.
We focus on practical aspects of the solver and regularization that are omitted from the main paper for brevity. The solver is GPU-accelerated using CuPy.

We solve \Cref{eq:costFunc} using scaled ADMM~\cite{ADMM}. Introducing a split variable $\mathbf{z}$ for the regularizer, the iterations are:
\begin{align}
(A^\top A + \rho I)\, v^{t+1} &= A^\top y + \rho\, (z^t - u^t), \label{eq:admm-v} \\[4pt]
z^{t+1} &= \operatorname{prox}_{(\lambda/\rho)\,\mathrm{TV}_{\mathbf{w}}}
\!\left(v^{t+1} + u^t\right), \label{eq:admm-z} \\[4pt]
z^{t+1} &\leftarrow \max\!\big(z^{t+1}, 0\big), \label{eq:admm-nn} \\[4pt]
u^{t+1} &= u^t + v^{t+1} - z^{t+1}. \label{eq:admm-u}
\end{align}
Here, $v$ is the data-fidelity variable, $z$ is the regularized variable, and $u$ is the scaled dual variable enforcing consensus.
The projection in \eqref{eq:admm-nn} enforces elementwise non-negativity.
We solve \eqref{eq:admm-v} inexactly using conjugate gradients, since $A$ includes spatial masking, making $A^\top A$ non-shift-invariant and therefore not diagonalizable in the Fourier domain.

\paragraph{Weighted anisotropic TV prior.}
We regularize $\mathbf{x}$ using a Haar-based anisotropic total variation (TV) prior implemented in the publicly available SpectralDiffuserCam codebase~\cite{diffSpec}.
Within the ADMM framework, the TV proximal operator is parameterized by a global regularization strength $\lambda/\rho$ and a scalar anisotropy parameter $\lambda_w$, which controls the relative strength of TV regularization across different tensor dimensions.

Specifically, let $a$ index the axes along which TV differences are computed.
We assign a fixed weight $w_a$ to each axis according to
\begin{equation}
w_a =
\begin{cases}
1, & a \in \{h,w\} \quad \text{(spatial dimensions)},\\
\lambda_w, & a = c \quad \text{(color dimension)},\\
\lambda_w/10, & a = p \quad \text{(polarization dimension)}.
\end{cases}
\label{eq:tv_weights}
\end{equation}
These weights remain fixed throughout optimization and are used internally by the TV proximal operator.

\paragraph{Reconstruction parameters.}
For real-data reconstructions using the physical polarization mask, we use
$\rho = 21$, $\lambda = 5 \times 10^{-1}$, $\lambda_w = 5 \times 10^{-1}$,
$50$ ADMM iterations, and a conjugate-gradient (CG) tolerance of $10^{-4}$
with a maximum of $100$ inner iterations.

For the matched-mask and controlled mismatch experiments in simulation, we use
the same optimization settings except for a reduced TV regularization strength,
which is optimal under matched forward-model conditions:
$\lambda = 5 \times 10^{-4}$ and $\lambda_w = 5 \times 10^{-4}$.
All other parameters are identical.

For lensless reconstructions obtained without a polarization mask, used as a
reference for evaluating mask-induced mismatch, we use
$\rho = 1$, $\lambda = 5 \times 10^{-1}$, $\lambda_w = 5 \times 10^{-1}$,
$50$ ADMM iterations, and a CG tolerance of $10^{-4}$ with $100$ inner iterations.
In this case, reconstruction is performed using $3$D TV (spatial and color
dimensions), rather than the $4$D TV used for polarization reconstructions.

\begin{figure}[htbp]
\def\resImgSz{0.16}
\centering
\begin{tabular}{c c c c}
\multicolumn{4}{c}{\textbf{Reference lensless reconstructions (polarization angle $0^\circ$ - $I_0$)}} \\[2pt]
\multicolumn{4}{c}{
\begin{tabular}{c c c}
No-mask & Matched & Measured \\[2pt]
\includegraphics[width=\resImgSz\columnwidth]{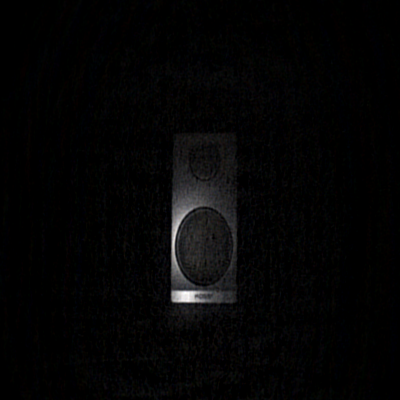} &
\includegraphics[width=\resImgSz\columnwidth]{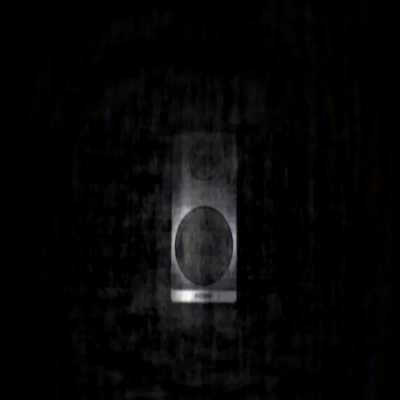} &
\includegraphics[width=\resImgSz\columnwidth]{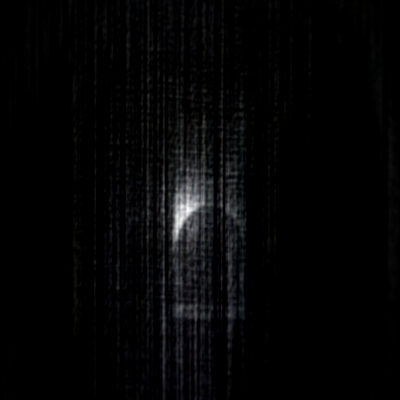}
\end{tabular}
} \\[5pt]

\multicolumn{4}{c}{\textbf{Controlled polarization mask mismatch}} \\[2pt]
Blur ($\sigma{=}0.5$) & Blur ($\sigma{=}3$) & Noise ($\sigma{=}0.02$) & Interpolated \\[2pt]
\includegraphics[width=\resImgSz\columnwidth]{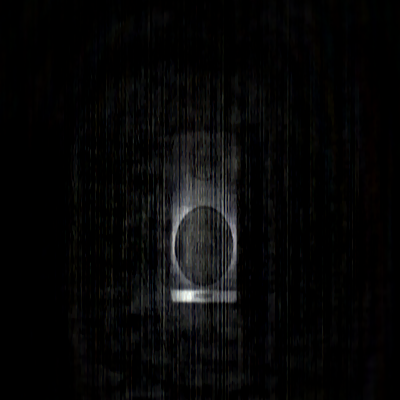} &
\includegraphics[width=\resImgSz\columnwidth]{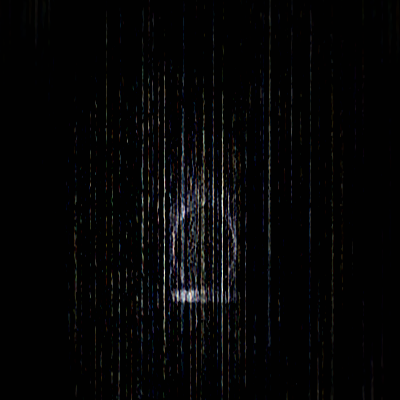} &
\includegraphics[width=\resImgSz\columnwidth]{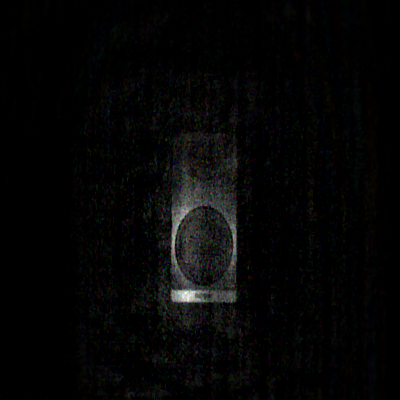} &
\includegraphics[width=\resImgSz\columnwidth]{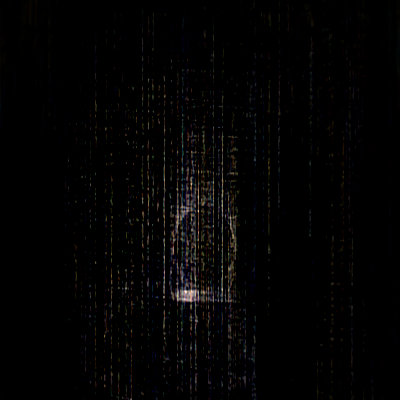} \\
\end{tabular}
\caption{\textbf{Effect of polarization mask mismatch on reconstruction quality.}
Top row: reference lensless reconstructions for the $I_{0}$ polarization channel obtained
(i) without a polarization mask (using a diffuser and external polarizer),
(ii) under a matched-mask setting in which the measured polarization mask is used consistently for both data generation and reconstruction, and
(iii) using the physical polarization mask in the prototype.
Bottom row: reconstructions obtained under controlled polarization-mask mismatch (1D spatial mask blur, additive mask noise, and interpolation between the ideal and measured masks), following the protocol described in \Cref{subsec:mismatch} of the main paper.}
\label{fig:mask-mismatch}
\end{figure}

\section{Additional Model Mismatch Analysis Results}
\label{sec:appendix_mismatch}

This section presents additional qualitative results that illustrate the effect of polarization mask mismatch on lensless polarization reconstruction, complementing the controlled mismatch analysis described in \Cref{subsec:mismatch} of the main paper.
All reconstructions are performed using the same measured diffuser PSF and identical reconstruction parameters.
For clarity, \Cref{fig:mask-mismatch} shows representative reconstructions for a single polarization channel ($I_{0}$) where similar trends are observed across the remaining polarization channels and scenes. 
Compared to the matched-mask reference, controlled polarization-mask mismatch produces characteristic reconstruction artifacts and contrast degradation that closely resemble those observed when using the physical prototype. 
This qualitative agreement suggests that the dominant degradation observed in real acquisitions is consistent with degradation expected from forward-model mismatch arising from deviations in the effective polarization mask response.

\section{Analysis of the effect of the distance between the polarization mask and the sensor}
\label{sec:appendix_mask_distance}

\begin{figure}[t]
    \centering
    \includegraphics[width=\linewidth]{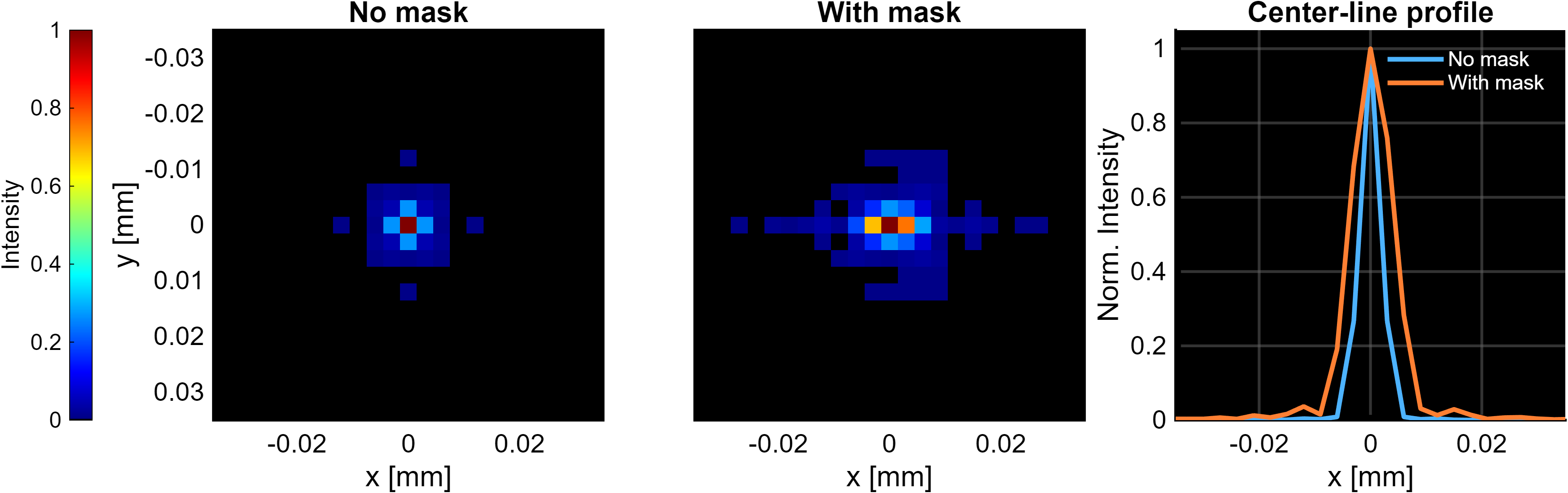}
    \caption{\textbf{Effect of mask-sensor distance on the system PSF.}
    (Left) Sensor-plane intensity without the polarization mask - a diffraction-limited PSF pattern.
    (Middle) Sensor-plane intensity with the  polarization mask placed $1.66\,\mathrm{mm}$ in front of the sensor, showing pronounced spreading perpendicular to the stripe direction.
    (Right) Center-line normalized intensity profiles extracted along the horizontal axis for both cases.}
    \label{fig:grating_effect}
\end{figure}

To justify the blur-based mismatch model used in \Cref{subsec:mismatch} of the main paper, we simulate the physical effect of the finite mask--sensor distance using scalar diffraction propagation implemented via the angular spectrum method.
The simulation is monochromatic with wavelength $\lambda=532\,\mathrm{nm}$ as a representative wavelength for the visible spectrum (small quantitative differences are expected for the red and blue shifts).
The optical field is sampled on a $4096\times4096$ grid with pixel pitch $\Delta x=3\,\mu\mathrm{m}$,which is chosen to be close to the physical sensor pixel pitch ($3.45\,\mu\mathrm{m}$) and to approximately match the sensor width. Therefore, the simulated plane spans $L=4096$ pixels $\times\,3\,\mu\mathrm{m}=12.288\,\mathrm{mm}$ per side.
Since explicitly simulating the diffuser is non-trivial, we replace it in this analysis by an ideal thin lens with a focal length of $f=20\,\mathrm{mm}$ and a circular pupil of radius $0.1L$ to generate a perfect converging spherical wave.
In such a configuration, the effects of the polarization filter are analyzed assuming a perfect lens imaging system. 
Under the assumption that the diffuser produces a superposition of many localized caustic features with delta-like structures, the effect of the polarization mask on each such local field can be conceptually extrapolated to the diffuser PSF by linear superposition.
The polarization mask is placed at a distance $z_2=d_{\mathrm{grating}}=1.66\,\mathrm{mm}$ in front of the sensor, corresponding to the combined thickness of the sensor cover glass ($\approx1.36\,\mathrm{mm}$) and the polarizer thickness ($\approx0.3\,\mathrm{mm}$, Thorlabs LPVISE2X2) in the laboratory prototype.
The field is propagated from the lens plane to the mask plane over $z_1=f-z_2$.
The polarization mask is modeled (for a single linear polarization orientation) as a one-dimensional amplitude grating  $T(x)$ consisting of vertical stripes of width $880\,\mu\mathrm{m}$ and four repeating amplitude transmission levels $\sqrt{\{0,\,0.5,\,1,\,0.5\}}$, as ideally fabricated.
Free-space propagation is implemented using the angular spectrum propagation method with transfer function
$H(f_x,f_y)=\exp\!\left(i k z \sqrt{1-(\lambda f_x)^2-(\lambda f_y)^2}\right)$,
while evanescent components are discarded.
Let $U_g(x,y)$ denote the complex optical field at the polarization mask plane for a given polarization, and let $T(x)$ denote the mask transmission.
The sensor-plane field after the mask is given by
\[
U_s(x,y)=\mathcal{P}_{z_2}\{T(x)\,U_g(x,y)\},
\]
where $\mathcal{P}_{z_2}\{\cdot\}$ denotes angular spectrum propagation over $z_2$.
For comparison, the field in the absence of the polarization mask is
\[
U_s^{(0)}(x,y)=\mathcal{P}_{z_2}\{U_g(x,y)\}.
\]
As shown in \Cref{fig:grating_effect}, inserting the polarization mask prior to propagation results in a slight one-dimensional spreading of the sensor-plane response along the horizontal direction (perpendicular to the stripe orientation, as expected from a 1D diffraction grating), compared to the reference propagation without the mask.
This observation indicates that the dominant prototype-specific effect induced by the finite distance between the polarization mask and the sensor is one-dimensional spatial mixing between adjacent polarization stripes.

\section{Additional Real-World Results}
\label{sec:appendix_additional_real}

This section presents additional qualitative results from real-world acquisitions captured with the physical lensless polarization prototype, complementing the representative examples shown in \Cref{subsec:real-world} of the main paper.
All scenes are reconstructed using the same forward model, reconstruction parameters, and measured diffuser PSF as detailed in the main paper.

\begin{figure}[t]
\newcommand{\rotateshift}{2.5cm}

\def\resImgSz{0.21}
\centering
\begin{tabular}{c c c c c}

        \rotatebox{90}{\parbox[c]{\rotateshift}{\centering prototype reconstruction  }} &

\includegraphics[width=\resImgSz\columnwidth]{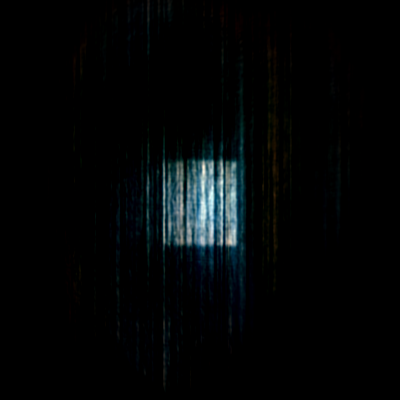} &
\includegraphics[width=\resImgSz\columnwidth]{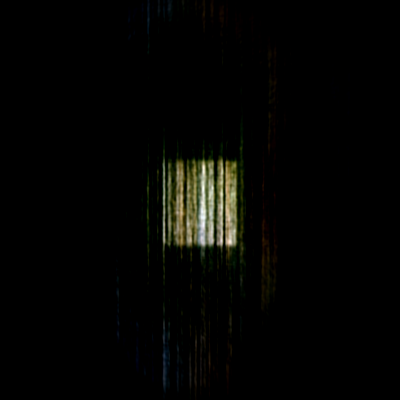} &
\includegraphics[width=\resImgSz\columnwidth]{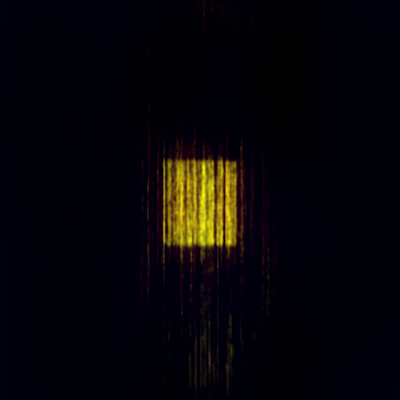} &
\includegraphics[width=\resImgSz\columnwidth]{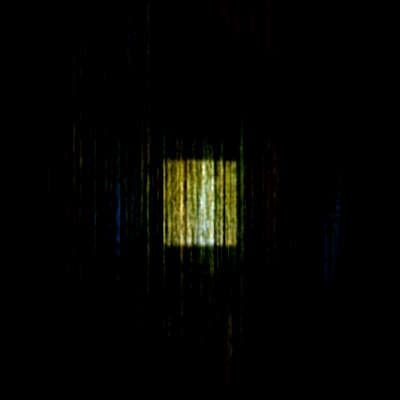} \\

\rotatebox{90}{\parbox[c]{\rotateshift}{\centering  matched-mask reconstruction  }} &
\includegraphics[width=\resImgSz\columnwidth]{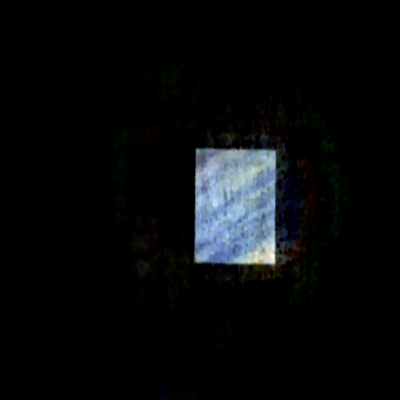} &
\includegraphics[width=\resImgSz\columnwidth]{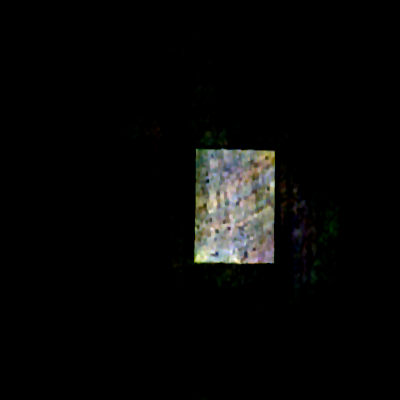} &
\includegraphics[width=\resImgSz\columnwidth]{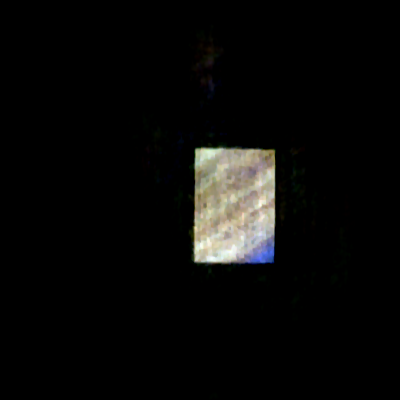} &
\includegraphics[width=\resImgSz\columnwidth]{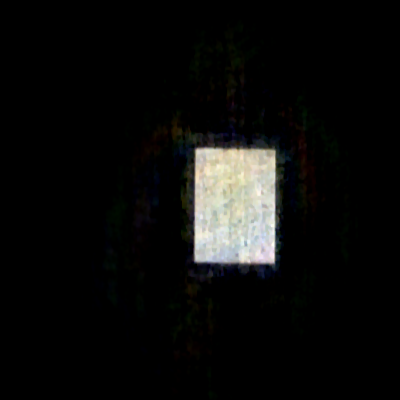} \\[1mm]

\small{(a) $I_{0}$} &
\small{(b) $I_{45}$} &
\small{(c) $I_{90}$} &
\small{(d) $I_{135}$} \\

\end{tabular}

\vspace{-1mm}
\caption{
\textbf{Back illumination examples}
For each scene, the first row shows the lensless polarization reconstruction using the physical prototype.
The second row shows the corresponding reconstruction in the matched-mask simulation setting, where the measured mask is used for simulation and reconstruction.
Columns correspond to the polarization sub-images
\(I_{0}\), \(I_{45}\), \(I_{90}\), and \(I_{135}\).
}
\label{fig:sim-back-ill-appendix}
\vspace{-2mm}
\end{figure}

\begin{figure}[t]
\def\resImgSz{0.21}
\centering
\begin{tabular}{c c c c}

\includegraphics[width=\resImgSz\columnwidth]{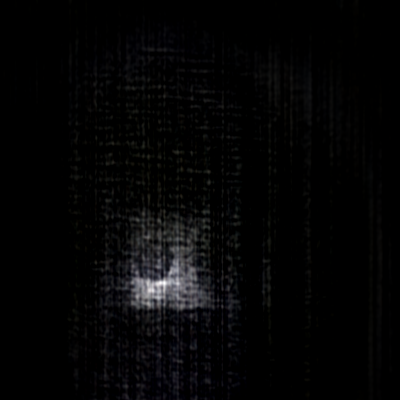} &
\includegraphics[width=\resImgSz\columnwidth]{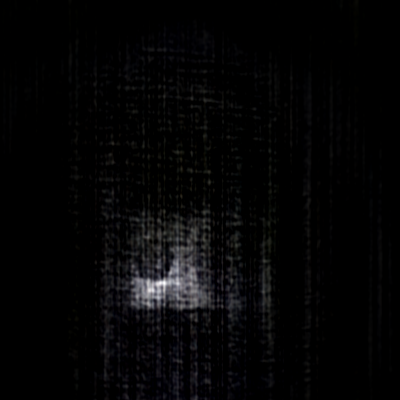} &
\includegraphics[width=\resImgSz\columnwidth]{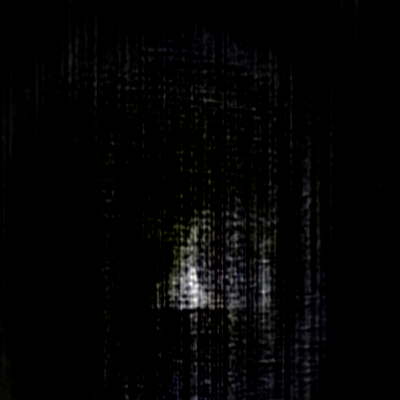} &
\includegraphics[width=\resImgSz\columnwidth]{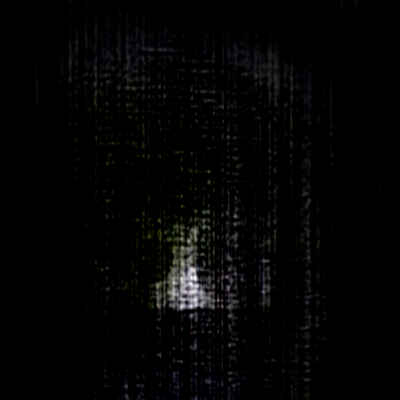} \\

\includegraphics[width=\resImgSz\columnwidth]{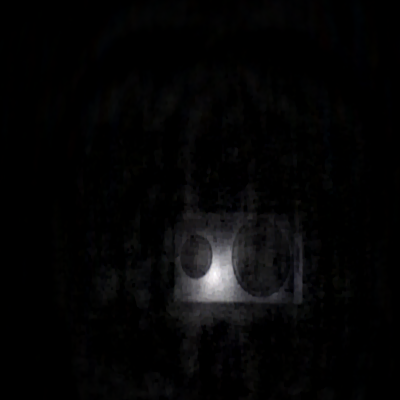} &
\includegraphics[width=\resImgSz\columnwidth]{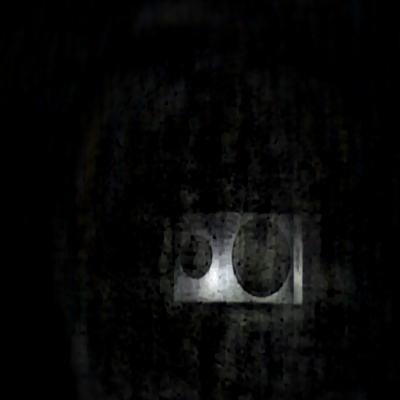} &
\includegraphics[width=\resImgSz\columnwidth]{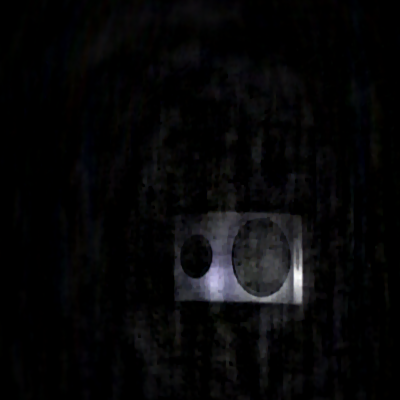} &
\includegraphics[width=\resImgSz\columnwidth]{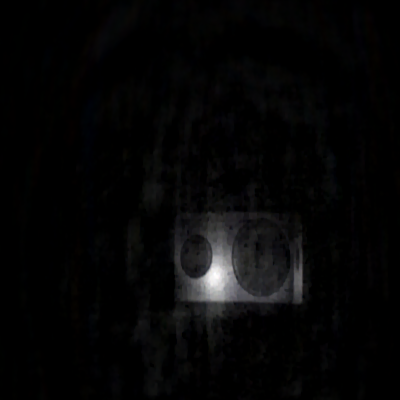} \\[1mm]

\includegraphics[width=\resImgSz\columnwidth]{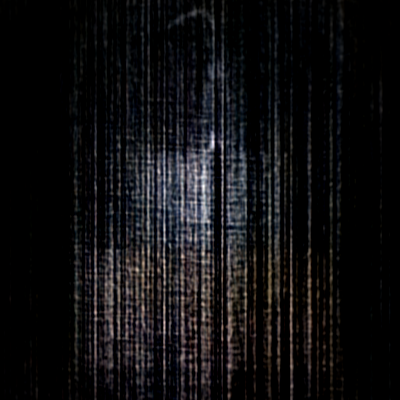} &
\includegraphics[width=\resImgSz\columnwidth]{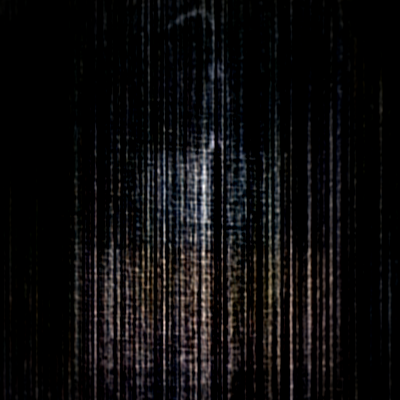} &
\includegraphics[width=\resImgSz\columnwidth]{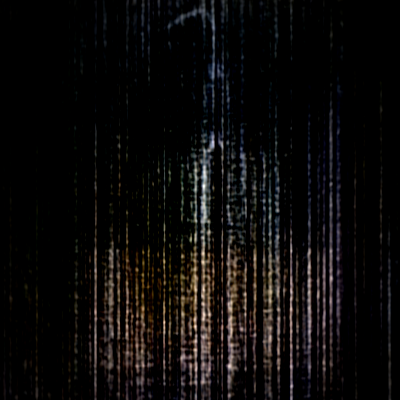} &
\includegraphics[width=\resImgSz\columnwidth]{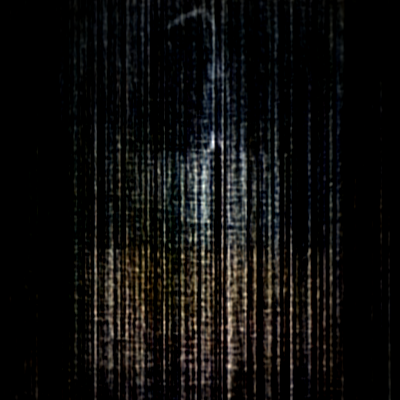} \\

\includegraphics[width=\resImgSz\columnwidth]{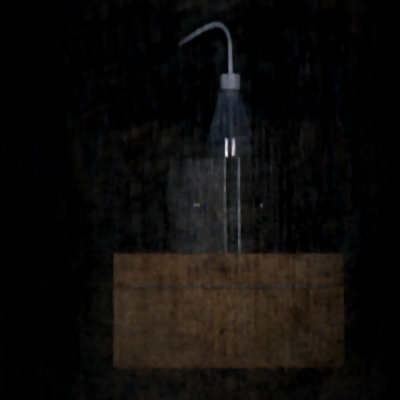} &
\includegraphics[width=\resImgSz\columnwidth]{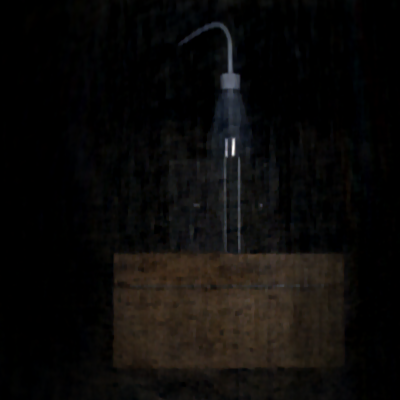} &
\includegraphics[width=\resImgSz\columnwidth]{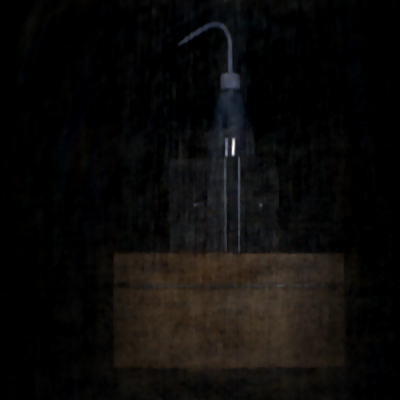} &
\includegraphics[width=\resImgSz\columnwidth]{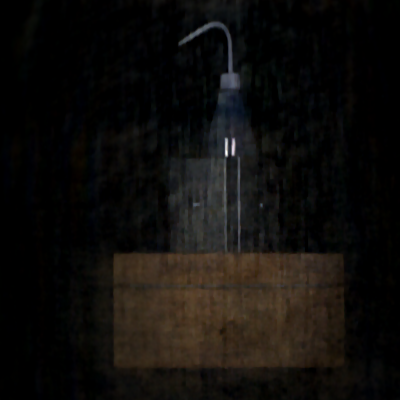} \\[1mm]

\includegraphics[width=\resImgSz\columnwidth]{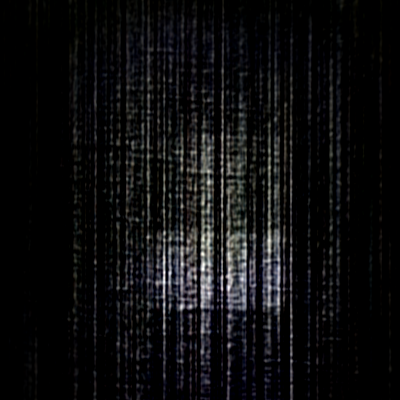} &
\includegraphics[width=\resImgSz\columnwidth]{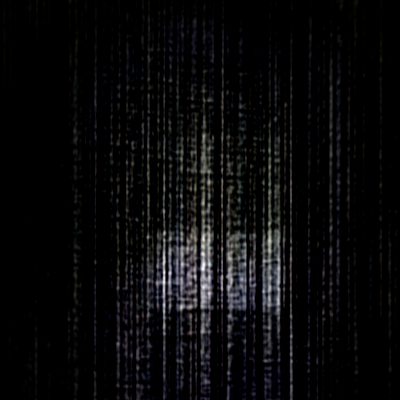} &
\includegraphics[width=\resImgSz\columnwidth]{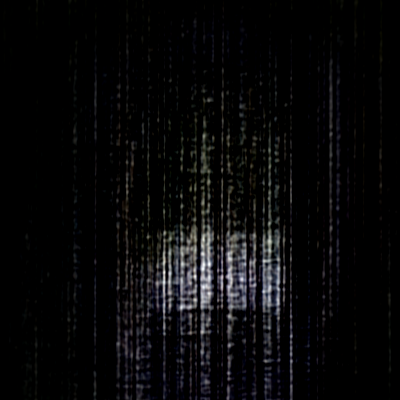} &
\includegraphics[width=\resImgSz\columnwidth]{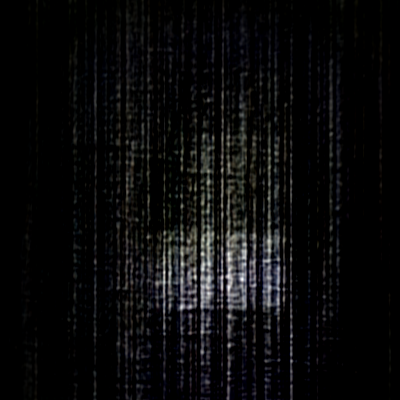} \\

\includegraphics[width=\resImgSz\columnwidth]{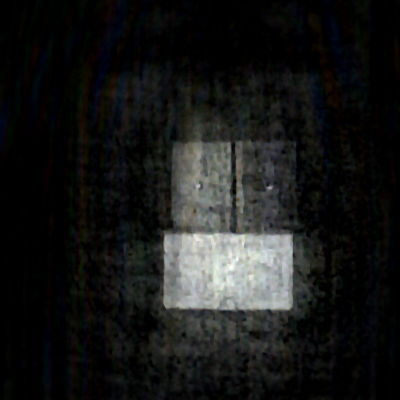} &
\includegraphics[width=\resImgSz\columnwidth]{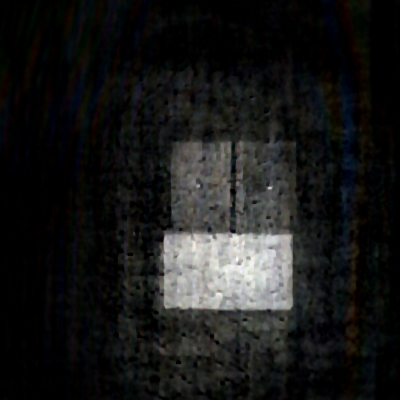} &
\includegraphics[width=\resImgSz\columnwidth]{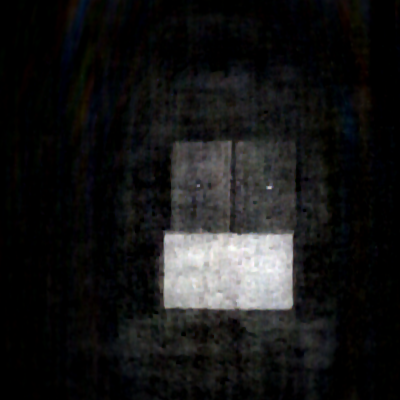} &
\includegraphics[width=\resImgSz\columnwidth]{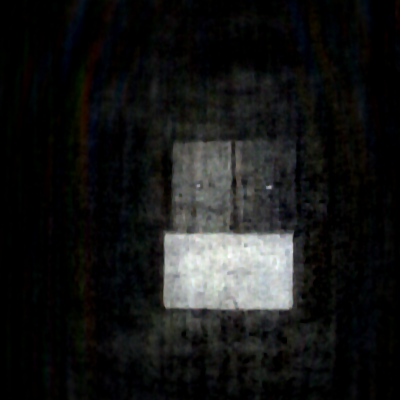} \\

\small{(a) $I_{0}$} &
\small{(b) $I_{45}$} &
\small{(c) $I_{90}$} &
\small{(d) $I_{135}$} \\

\end{tabular}

\vspace{-1mm}
\caption{
\textbf{Front illumination examples}
For each scene, the first row shows the lensless polarization reconstruction using the physical prototype.
The second row shows the corresponding reconstruction in the matched-mask simulation setting, where the measured mask is used for simulation and reconstruction.
Columns correspond to the polarization sub-images
\(I_{0}\), \(I_{45}\), \(I_{90}\), and \(I_{135}\).
}
\label{fig:sim-front-ill-appendix}
\vspace{-2mm}
\end{figure}

For each scene, we show polarization reconstructions obtained from a single physical snapshot alongside the corresponding matched-mask simulation reconstruction, generated by synthetically multiplexing sequential single-polarization lensless measurements using the measured polarization mask in the same protocol as detailed in \Cref{subsec:matched} of the main paper.
Due to the need to remove and reinstall the physical polarization mask between acquisitions, the physical and simulated scenes are similar but not identical, and may exhibit spatial misalignments.
The comparisons in \Cref{fig:sim-front-ill-appendix,fig:sim-back-ill-appendix} are therefore qualitative.

Across all scenes, the physical reconstructions recover the dominant scene structure and polarization-dependent illumination trends, including consistent polarization-dependent differences between the $I_{0}$/$I_{90}$ and $I_{45}$/$I_{135}$ channels.
Compared to the matched-mask simulation setting, the real reconstructions exhibit reduced contrast and increased spatial artifacts, with attenuated polarization cues.

These degradation patterns are consistent across scenes and closely resemble the effects induced by controlled polarization-mask mismatch in the main paper (\Cref{subsec:mismatch}), supporting mask-related mismatch as a primary contributor to the observed sim–real performance gap.
All results shown represent the reconstruction quality obtained with the current prototype under the settings used throughout this work.

\end{document}